\documentclass[]{emulateapj}
\usepackage{natbib,amsmath}

\begin{document}

\newcommand{\noprint}[1]{}
\newcommand{\figsetstart}{{\bf Fig. Set} }
\newcommand{\figsetend}{}
\newcommand{\figsetgrpstart}{}
\newcommand{\figsetgrpend}{}
\newcommand{\figsetnum}[1]{{\bf #1.}}
\newcommand{\figsettitle}[1]{ {\bf #1} }
\newcommand{\figsetgrpnum}[1]{\noprint{#1}}
\newcommand{\figsetgrptitle}[1]{\noprint{#1}}
\newcommand{\figsetplot}[1]{\noprint{#1}}
\newcommand{\figsetgrpnote}[1]{\noprint{#1}}

\def\nhdlls{157}
\def\ntot{16X}
\def\nslls{99}
\def\hub{h_{72}^{-1}}
\def\umfp{{\hub \, \rm Mpc}}
\def\mew{W_\lambda}
\def\ew{$\mew$}
\def\mzq{z_q}
\def\mfmin{f_{\rm min}}
\def\fmin{$\mfmin$}
\def\zabs{$z_{\rm abs}$}
\def\mzabs{z_{\rm abs}}
\def\msna{{\rm S/N}^{\rm A}_{912}}
\def\sna{S/N$^{\rm A}_{912}$}
\def\mnull{\nu_{\rm 912}}
\def\nnull{$\nu_{\rm 912}$}
\def\intl{\int\limits}
\def\nstatqso{193}
\def\maxoff{0.4}
\def\clls{1.9 \pm 0.2}
\def\alls{5.2 \pm 1.5}
\def\blls{-0.9^{+0.4}_{-0.05}}
\def\cmma{\;\;\; ,}
\def\perd{\;\;\; .}
\def\ltk{\left [ \,}
\def\ltp{\left ( \,}
\def\ltb{\left \{ \,}
\def\rtk{\, \right  ] }
\def\rtp{\, \right  ) }
\def\rtb{\, \right \} }
\def\sci#1{{\; \times \; 10^{#1}}}
\def \rAA {\rm \AA}
\def \zem {$z_{\rm em}$}
\def \mzem {z_{\rm em}}
\def \mzlls {z_{\rm LLS}}
\def \zlls {$z_{\rm LLS}$}
\def \mzpeak {\mzlls^{\rm peak}}
\def \zpeak {$\mzpeak$}
\def \mzend {z_{\rm end}}
\def \zend {$z_{\rm end}$}
\def \mzstrtt {z_{\rm start}^{\rm S/N^A=2}}
\def \zstrtt {$z_{\rm start}^{\rm S/N^A=2}$}
\def \zstrto {$z_{\rm start}^{\rm S/N^A=1}$}
\def \mzstrto {z_{\rm start}^{\rm S/N^A=1}}
\def \zstrt {$z_{\rm start}$}
\def \mzstrt {z_{\rm start}}
\def\smm{\sum\limits}
\def \lll  {$\lambda_{\rm 912}$}
\def \mlll  {\lambda_{\rm 912}}
\def \mtll  {\tau_{\rm 912}}
\def \tll  {$\tau_{\rm 912}$}
\def \mavtll  {<\mtll>}
\def \tigm  {$\tau_{\rm IGM}$}
\def \mtigm  {\tau_{\rm IGM}}
\def \cmm  {cm$^{-2}$}
\def \cmmm {cm$^{-3}$}
\def \kms  {km~s$^{-1}$}
\def \mkms  {{\rm km~s^{-1}}}
\def \lyaf {Ly$\alpha$ forest}
\def \Lya  {Ly$\alpha$}
\def \lya  {Ly$\alpha$}
\def \mlya  {Ly\alpha}
\def \Lyb  {Ly$\beta$}
\def \lyb  {Ly$\beta$}
\def \lyg  {Ly$\gamma$}
\def \ly5  {Ly-5}
\def \ly6  {Ly-6}
\def \ly7  {Ly-7}
\def \nhi  {$N_{\rm HI}$}
\def \mnhi  {N_{\rm HI}}
\def \lnhi {$\log N_{HI}$}
\def \mlnhi {\log N_{HI}}
\def \etal {\textit{et al.}}
\def \ob {$\Omega_b$}
\def \obh {$\Omega_bh^{-2}$}
\def \om {$\Omega_m$}
\def \ol {$\Omega_{\Lambda}$}
\def \gz {$g(z)$}
\def \mgz {g(z)}
\def \lyaf {Lyman--$\alpha$ forest}
\def \fnhi {$f(\mnhi,X)$}
\def \mfnhi {f(\mnhi,X)}
\def \myfnhi {f_{\rm Ly\alpha}(\mnhi,X)}
\def \yfnhi {$f_{\rm Ly\alpha}(\mnhi,X)$}
\def \lfnhi {$f_{\rm LLS}(\mnhi,X)$}
\def \dfnhi {$f_{\rm DLA}(\mnhi,X)$}
\def \sfnhi {$f_{\rm SLLS}(\mnhi,X)$}
\def \mlfnhi {f_{\rm LLS}(\mnhi,X)}
\def \mdfnhi {f_{\rm DLA}(\mnhi,X)}
\def \msfnhi {f_{\rm SLLS}(\mnhi,X)}
\def \ztot {$\Delta z_{\rm TOT}$}
\def \mztot {\Delta z_{\rm TOT}}
\def \mlplls {\ell_{\rm{PLLS}}(z)}
\def \lplls {$\ell_{\rm{PLLS}}(z)$}
\def \mlzlls {\ell_{\rm{LLS}}(z)}
\def \lzlls {$\ell_{\rm{LLS}}(z)$}
\def \mllls {\ell_{\rm{LLS}}(X)}
\def \llls {$\ell_{\rm{LLS}}(X)$}
\def \ldla {$\ell_{\rm{DLA}}(X)$}
\def \lslls{$\ell_{\rm{SLLS}}(X)$}
\def \mlslls{\ell_{\rm{SLLS}}(X)}
\def \nlls {$n_{\rm LLS}$}
\def \slls {$\sigma_{\rm LLS}$}
\def \mnlls {n_{\rm LLS}}
\def \mslls {\sigma_{\rm LLS}}
\def \drlls {$\Delta r_{\rm LLS}$}
\def \mdrlls {\Delta r_{\rm LLS}}
\def \mlmfp {\lambda_{\rm mfp}^{912}}
\def \lmfp {$\lambda_{\rm mfp}^{912}$}
\def \mbplls {\beta_{\rm pLLS}}
\def \bplls {$\beta_{\rm pLLS}$}
\def \btlls {$\beta_{\rm LLS}$}
\def \mbtlls {\beta_{\rm LLS}}
\def \lteff {$\tau_{\rm eff,LL}$}
\def \teff {$\tau_{\rm eff,LL}$}
\def \tlya {$\tau_{\rm eff}^{\rm Ly\alpha}$}
\def \mtlya {\tau_{\rm eff}^{\rm Ly\alpha}}
\def \mteff {\tau_{\rm eff}}
\def \mllteff {\mteff^{912}}
\def \llteff {$\mteff^{912}$}
\def \mnmin {\mnhi^{\rm min}}
\def \nmin {$\mnhi^{\rm min}$}
\def \O {${\mathcal O}(N,X)$}
\newcommand{\cm}[1]{\, {\rm cm^{#1}}}
\def\N#1{{N({\rm #1})}}
\def\psol#1#2#3#4{$\{ {\rm #1}^{#2}/{\rm #3}^{#4}\}$}
\def\pxh{$\{ {\rm X/H} \}$}
\def \snrlim {SNR$_{lim}$}
\def\mglls {\gamma_{\rm LLS}}
\def\mavgt {<\mtll>}

\title{The Keck $+$ Magellan Survey for Lyman Limit Absorption III:  
Sample Definition and Column Density Measurements}

\author{
J. Xavier Prochaska\altaffilmark{1}, 
John M. O'Meara\altaffilmark{2}, 
Michele Fumagalli\altaffilmark{3,4},
Rebecca A. Bernstein\altaffilmark{4}, 
Scott M. Burles\altaffilmark{6}
}
\altaffiltext{1}{Department of Astronomy and Astrophysics, UCO/Lick
  Observatory, University of California, 1156 High Street, Santa Cruz,
  CA 95064, USA}
\altaffiltext{2}{Department of Chemistry and Physics, Saint Michael's College.
One Winooski Park, Colchester, VT 05439, USA}
\altaffiltext{3}{Institute for Computational Cosmology, Department of Physics, Durham University,
 South Road, Durham, DH1 3LE, UK}
\altaffiltext{4}{Observatories of the Carnegie Institution for Science, 813 Santa Barbara Street,
  Pasadena, CA 91101, USA}
\altaffiltext{6}{Cutler Group, LP., 101 Montgomery St., San Francisco,
  CA 94104, USA}

\begin{abstract}
We present an absorption-line survey of optically thick gas clouds
-- Lyman Limit Systems (LLSs) -- observed at high dispersion with spectrometers
on the Keck and Magellan telescopes.  We measure column
densities of neutral hydrogen \nhi\ and associated metal-line
transitions for \nhdlls~LLSs at 
$\mzlls = 1.76-4.39$ restricted to 
$10^{17.3} \cm{-2} \le \mnhi < 10^{20.3} \cm{-2}$.
An empirical analysis of ionic ratios indicates an increasing
ionization state of the gas with decreasing \nhi\ and that the
majority of LLSs are highly ionized, confirming 
previous expectations.  
The Si$^+$/H$^0$ ratio spans
nearly four orders-of-magnitude, implying a large dispersion in the
gas metallicity.  Fewer than 5\%\ of these LLSs have no
positive detection of a metal transition;
by $z \sim 3$, nearly all gas that is dense enough to exhibit a very high Lyman
limit opacity has previously been polluted by heavy elements.  
We add new measurements to the small subset of LLS ($\approx 5-10\%$) 
that may have super-solar abundances.
High Si$^+$/Fe$^+$ ratios suggest an $\alpha$-enhanced medium whereas
the Si$^+$/C$^+$ ratios do not exhibit the super-solar enhancement
inferred previously for the \lya\ forest.  
\end{abstract}


\section{Introduction}

As a packet of ionizing radiation ($h \nu \ge 1$\,Ryd) traverses the
universe, it has a high probability of encountering a slab of
optically thick, \ion{H}{1} gas.  For sources in the $z \sim 4$
universe the mean free path is only $\approx 30$\,Mpc
\citep[physical;][]{worseck+14}, i.e.\ less than 2\%\ of the
event horizon.   Observationally, researchers refer to this optically
thick gas as Lyman limit systems (LLSs) owing to their unmistakable
signature of continuum opacity at the Lyman limit ($\approx 912$\AA)
in the system restframe.
A fraction of this gas lies within the dense, neutral interstellar
medium (ISM) of galaxies, yet the majority of opacity must arise
from gas outside the ISM \citep[e.g.][]{fpk+11,ribaudo11}.
Indeed, the interplay between galaxies and the LLS is a highly active
area of research which includes studies of the so-called
circumgalactic medium 
\citep[CGM; e.g.][]{steidel+10,werk+13,QPQ7}.

For many decades, LLS have been surveyed in quasar
spectra \citep[e.g.][]{tytler82,ssb,storrie94}, albeit often from
heterogeneous samples.  These works established the high incidence
of LLSs which evolves rapidly with redshift.  With the realization of
massive spectral datasets, a renaissance of LLS surveys has followed
yielding statistically robust measurements from 
homogenous and well-selected quasar samples 
\citep{pow10,songaila10,ribaudo11,omeara13,fop+13}.
Analysis of these hundreds of systems reveals an incidence of
approximately 1.2 systems per unit redshift at $z \sim 3$ that evolves
steeply with redshift $\ell(z) \propto (1+z)^{1.5}$ for $z \approx
1-5$ \citep{ribaudo11,fop+13}.
With these same spectra,
researchers have further measured the mean free path of ionizing radiation
\citep[\lmfp;][]{pwo09,omeara13,fop+13,worseck+14}, which 
sets the intensity and shape of the extragalactic UV background. 
Following the redshift evolution of the LLS incidence, \lmfp\ also
evolves steeply with the expanding universe, implying a more highly
ionized universe with advancing cosmic time \citep{worseck+14}.

The preponderance of LLSs bespeaks a major reservoir of baryons.  In
particular, given the apparent paucity of heavy elements within
galaxies \cite[e.g.][]{bouche06II,peeples+14}, the LLSs may present the dominant
reservoir of metals in the universe \citep[e.g.][]{poh+06}. 
However, a precise calculation of the heavy elements within LLSs and their
contribution to the cosmic budget has not yet been achieved.
Despite our success at surveying hundreds of LLSs, there have been few
studies resolving their physical properties and these have generally
examined a few individual cases
\citep[e.g.][]{steidel90,pro99} or composite spectra \citep{fop+13}.
This reflects both the challenges related to data acquisition and
analysis together with a historical focus in the community towards the
ISM of galaxies (probed by DLAs) 
and the more diffuse intergalactic medium (IGM).

At $z>2$, a few works 
have examined the set of LLSs with high \ion{H}{1} column density
($\mnhi \ge 10^{19} \cm{-2}$),  generally termed the
super-LLSs or sub-damped \lya\ systems.  Their \nhi\ frequency
distribution \fnhi\ and chemical abundances have been analyzed from 
a modestly sized sample \citep{mirka03,peroux05,opb+07,zafar13,som13}.  
Ignoring ionization corrections, which may not be justified,
these SLLSs exhibit metallicities of approximately 1/10 solar,
comparable to the enrichment level of the higher-\nhi, damped \lya\
systems \citep[DLAs;][]{rafelski+12}.
In addition, a few LLSs have received special attention owing
to their peculiar metal-enrichment \citep{poh+06,fop11} and/or the
detection of D for studies of Big Bang Nucleosynthesis
\citep[e.g.][]{bt_1937,obp+06}. 
Most recently, a sample of 15 LLSs has been surveyed for highly ionized
\ion{O}{6} absorption \citep{lehner+14}, which is present at a high
rate.  A comprehensive study of the absorption-line properties of the
LLSs at high redshift, however, has not yet been performed.

Scientifically, we have two primary motivations to survey the LLSs at
$z>2$.  First and foremost, we aim to dissect the physical nature of
the gas that dominates the opacity to ionizing radiation in the
universe.  One suspects that these LLSs trace a diverse set of
overdense structures ranging from galactic gas to the densest
filaments of the
cosmic web.   Such diversity may manifest in an wide distribution of
observed properties (e.g.\ metal enrichment, ionization state,
kinematics).  
Second, modern theories of galaxy formation predict that the gas
fueling star formation accretes onto galaxies in cool, dense streams
\citep[e.g.][]{kkw+05,dbe+08}.  Radiative transfer analysis of
hydrodynamic simulations of this process predict a relatively high
cross-section of optically thick gas around galaxies
\citep[e.g.][]{fg11,fpk+11,fhp+14,fg+14}.  
Indeed, an optically thick CGM
envelops the massive galaxies hosting $z \sim 2$ quasars
\citep{QPQ2,qpq6}, LLSs are observed near Lyman break galaxies 
\citep{rudie12}, and such gas
persists around present-day $L^*$ galaxies 
\citep[e.g.][]{chg+10,werk+13}.  
The latter has inspired, in part, surveys of the LLSs at $z<1$ with 
ultraviolet spectroscopy \citep[e.g.][]{ribaudo11,lht+13}.


Thus motivated, we have obtained a large dataset of high-dispersion
spectroscopy on $z>3$ quasars at the Keck and Las Campanas
Observatories.  We have supplemented this program with additional
spectra obtained to survey the damped \lya\ systems
\citep[e.g.][]{pwh+07,berg+15a} and the intergalactic medium
\citep[e.g.][]{fpl+08}.  In this paper, we present the
comprehensive dataset of column density measurements on over 150~LLSs.
Future manuscripts will examine the metallicity, chemical abundances,
kinematics, and ionization state of this gas.
This manuscript is outlined as follows:
Section~\ref{sec:data} describes the dataset analyzed including a
summary of the observations and procedures for generating calibrated
spectra.  We define an LLS in Section~\ref{sec:define} and detail the
procedures followed to estimate the \ion{H}{1} column densities in
Section~\ref{sec:NHI}. 
Section~\ref{sec:ionic} presents measurements of the ionic column
densities and the primary results of an empirical assessment of these
data are given in Section~\ref{sec:results}.   A summary 
in Section~\ref{sec:summary} concludes the paper.  

\section{Data}
\label{sec:data}

This section describes the steps taken to generate a large dataset of
high-dispersion, calibrated spectra of high redshift LLSs.

\subsection{Our Survey}
\label{sec:sample}

The sample presented in this manuscript is intended to be a nearly,
all-inclusive set of LLSs discovered in the high-dispersion (echelle or
echellette; $R>5,000$) spectra that we have gathered at the Keck and
Magellan telescopes.  Regarding Keck, we have examined all of the data
obtained by Principal Investigators (PIs) A.M. Wolfe and J.X. Prochaska at
the W.M. Keck Observatory through April 2012, and from PIs
Burles, O'Meara, Bernstein, and Fumagalli at Magellan through July 2012.
We also include the Keck spectra analyzed by \cite{pps+10}.

Each spectrum was visually inspected for the presence of
damped \lya\ absorption and/or a continuum break at wavelengths
$\lambda < 912$\AA\ in the quasar rest-frame.
The complex combination of spectral S/N, wavelength
coverage, and quasar emission redshift \zem\ leads to a varying
sensitivity to an LLS.  No attempt is made here to define a statistical
sample, e.g. to assess the random incidence of LLSs nor their \nhi\
frequency distribution \fnhi.  We refer the reader to previous
manuscripts on this topic \citep{pow10,fop+13}. 
Because our selection is based solely on \ion{H}{1} absorption, however, we believe
the sample is largely
unbiased with respect to other properties of the gas,
e.g.\ metal-line absorption, kinematics, ionization state.

The sample was limited during the survey by:
 (1) generally ignoring LLSs with absorption redshifts within
 3000\,\kms\ of the reported quasar redshift
\zem, so-called proximate LLS or PLLS;
and
(2) generally ignoring LLSs with $\mnhi <
10^{17.3} \cm{-2}$, especially when the S/N was poor near the Lyman limit.
We note further that many of the Keck spectra were obtained to study
damped \lya\ systems (DLAs) at $z >2$
\citep[e.g.][]{pro01,pwh+07,rafelski+12,marcel13,berg+15a}.  We have ignored
systems targeted as DLAs and also absorbers within $\approx
1500 \mkms$ of these DLAs because the DLA system complicates analysis
of the \ion{H}{1} Lyman series and metal-line transitions of any
nearby LLS.
In $\S$~\ref{sec:define}, we offer a strict definition for an LLS to
define our sample of \nhdlls\ systems.

\begin{deluxetable*}{lcccccccccc}
\tablewidth{0pc}
\tablecaption{JOURNAL OF HIRES OBSERVATIONS\label{tab:hires}}
\tabletypesize{\scriptsize}
\tablehead{
\colhead{QSO} & \colhead{Alt. Name} & \colhead{RA} & \colhead{DEC} &
\colhead{$r/V^a$} & \colhead{$z_{em}$} & \colhead{Date} 
& \colhead{Slit$^b$} 
& \colhead{Mode} 
& \colhead{Exp} 
& \colhead{S/N$^c$} \\ 
&& (J2000) & (J2000) & (mag) &     & (UT)  & & & (s) & (pix$^{-1}$)}
\startdata
SDSS0121+1448 &                & 01:21:56.03& +14:48:23.8 & 17.1  & 2.87  & 08 Sep 2004  & C1 & HIRESb & 7200  & 15/26\\
PSS0133+0400  &                & 01:33:40.4 & +04:00:59   &  18.3     & 4.13  & 27 Dec 2006  & C1 & HIRESr & 7200  & 14/20 \\
SDSS0157-0106 &                & 01:57:41.56& -01:06:29.6 & 18.2  & 3.564 & 18 Dec 2003  & C5 & HIRESr & 9000  & X/14\\
Q0201+36      &                & 02:04:55.60& +36:49:18.0 & 17.5  & 2.912 & 06 Oct 2004  & C1 & HIRESb & 3600  & 4.5/9\\
PSS0209+0517  &                & 02:09:44.7 & +05:17:14   & 17.8  & 4.18  & 18 Sep 2007  & C1 & HIRESr & 11100 & 31/24 \\
Q0207--003    &                & 02:09:51.1 & --00:05:13  & 17.1  & 2.86  & 08 Sep 2004  & C1 & HIRESb & 5400  & 15/40\\
              &                &            &             &       &       & 09 Sep 2004  & C1 & HIRESb & 8100  & \\
LB0256--0000  &                & 02:59:05.6 & +00:11:22   & 17.7  & 3.37  & 03 Jan 2006  & C5 & HIRESb & 7049   & 11/17  \\
Q0301--005 &                & 03:03:41.0 & --00:23:22  & 17.6  & 3.23  & 09 Sep 2004  & C1 & HIRESb & 7800  & X/15 \\
Q0336--01     &                & 03:39:01.0 & --01:33:18  & 18.2  & 3.20  & 26 Oct 2005  & C5 & HIRESb & 3600  & X/10 \\
              &                &            &             &       &       & 01 Nov 2003  & C1 & HIRESr & 10800 & 15 \\
SDSS0340-0159 &                & 03:40:24.57& --05:19:09.2& 17.95 & 2.34  & 06 Oct 2008  & C1 & HIRESb & 3000  & 7/15 \\
HE0340-2612   &                & 03:42:27.8 & --26:02:43  & 17.4  & 3.14  & 26 Oct 2005  & C1 & HIRESb & 7200  & 17/X    \\ 
SDSS0731+2854 &                & 07:31:49.5 &  +28:54:48.6& 18.5  & 3.676 & 04 Jan 2006  & C5 & HIRESb & 7200  & X/15   \\
Q0731+65      &                & 07:36:21.1 &  +65:13:12  & 18.5  & 3.03  & 28 Oct 2005  & C5 & HIRESb & 5400  & X/16 \\
              &                &            &             &       &       & 04 Jan 2006  & C5 & HIRESb & 7200  & X/12\\
J0753+4231    &                & 07:53:03.3 &  +42:31:30  & 17.92 & 3.59  & 26 Oct 2005  & C5 & HIRESb & 3300  & X/12\\
              &                &            &             &       &       & 28 Oct 2005  & C5 & HIRESb & 4800  & X/16\\
SDSS0826+3148 &                & 08:26:19.7 & +31:48:48   & 17.76 & 3.093 & 27 Dec 2006  & C1 & HIRESr & 7900  & 37/22 \\
J0828+0858    &                & 08:28:49.2 & +08:58:55   & 18.30 & 2.271 & 14 Apr 2012  & C1 & HIRESb & 1295  & 6/9 \\
J0900+4215    &                & 09:00:33.5 & +42:15:46   & 16.98 & 3.290 & 15 Apr 2005  & C1 & HIRESb & 4700  & X/20 \\
J0927+5621    &                & 09:27:05.9 & +56:21:14   & 18.22 & 2.28  & 14 Apr 2005  & C5 & HIRESb & 8500  & 6/20 \\
J0942+0422    &                & 09:42:02.0 & +04:22:44   & 17.18 & 3.28  & 18 Mar 2005  & C1 & HIRESb & 7200  & 27/X\\
J0953+5230    &                & 09:53:09.0 & +52:30:30   & 17.66 & 1.88  & 18 Mar 2005  & C1 & HIRESb & 7200  & 18/22 \\
Q0956+122     &                & 09:58:52.2 & +12:02:44   & 17.6  & 3.29  & 03 Jan 2006  & C5 & HIRESb & 7200  & X/40  \\
              &                &            &             &       &       & 07 Apr 2006  & C1 & HIRESr & 1800  & 15/10 \\
HS1011+4315   &                & 10:14:47.1 & +43:00:31   & 16.1  & 3.1   & 14 Apr 2005  & C5 & HIRESb & 5100  & X/40 \\
              &                &            &             &       &       & 27 Apr 2007  & B2 & HIRESr & 3600  & 47/47 \\
              &                &            &             &       &       & 28 Apr 2007  & B2 & HIRESr & 3600  & 47/47 \\
J1019+5246    &                & 10:19:39.1 & +52:46:28   & 17.92 & 2.170 & 11 Apr 2007  & C1 & HIRESb & 7200  & 11/16 \\
Q1017+109     &                & 10:20:10.0 & +10:40:02   & 17.5  & 3.15  & 06 Apr 2006  & C5 & HIRESb & 7200  &  25/X \\
J1035+5440    &                & 10:35:14.2 & +54:40:40   & 18.21 & 2.988 & 25 Mar 2008  & C1 & HIERSr & 10800 & 23/24 \\
SDSS1040+5724 &                & 10:40:18.5 & +57:24:48   & 18.30 & 3.409 & 04 Jan 2006  & C5 & HIRESb & 8100  & X/12 \\
Q1108-0747    &                & 11:11:13.6 & --08:04:02  & 18.1  & 3.92  & 07 Apr 2006  & C1 & HIRESr & 7200  & 30/10 \\
J1131+6044    &                & 11:31:30.4 & +60:44:21   & 17.73 & 2.921 & 26 Dec 2006  & C1 & HIRESb & 7200  & 14/18 \\
J1134+5742    &                & 11:34:19.0 & +57:42:05   & 18.20 & 3.522 & 05 Jan 2006  & C5 & HIRESr & 6300  & 26/22 \\
J1159-0032    &                & 11:59:40.7 & --00:32:03  & 18.10 & 2.034 & 14 Apr 2012  & C1 & HIRESb & 2400  & 5/7 \\
Q1206+1155    &                & 12:09:18.0 & +09:54:27   & 17.6  & 3.11  & 06 Apr 2006  & C5 & HIRESb & 7200  & 23/X \\
Q1330+0108    &                & 13:32:54.4 & +00:52:51   & 18.2  & 3.51  & 07 Apr 2006  & C1 & HIRESr & 7200  & 11/9 \\
HS1345+2832   &                & 13:48:11.7 & +28:18:02   & 16.8  & 2.97  & 14 Apr 2005  & C5 & HIRESb & 4800  & X/27\\
PKS1354--17     &                 & 13:57:06.07 & --17:44:01.9 & 18.5 & 3.15 & 28 Apr 2007& C5 &  HIRESr & 7200 & 8/7 \\
J1407+6454    &                & 14:07:47.2 & +64:54:19   & 17.24 & 3.11  & 14 Apr 2005  & C5 & HIRESb & 5400  & X/20 \\
HS1431+3144   &                & 14:33:16.0 & +31:31:26   & 17.1  & 2.94  & 06 Apr 2006  & C5 & HIRESb & 6000  & 25/43 \\
J1454+5114    &                & 14:54:08.9 & +51:14:44   & 17.59 & 3.644 & 14 Jul 2005  & C5 & HIRESr & 1800  & 10/7.5 \\
J1509+1113    &                & 15:09:32.1 & +11:13:14   & 19.0  & 2.11  & 15 Apr 2012  & C1 & HIRESb & 5200  & 4/7 \\
J1555+4800    &                & 15:55:56.9 & +48:00:15   & 19.1  & 3.297 & 15 Apr 2005  & C5 & HIRESr & 10800 & 13/10 \\
              &                &            &             &       &       & 14 Jul 2005  & C5 & HIRESr & 10800 &       \\
              &                &            &             &       &       & 04 Jun 2006  & C5 & HIRESr & 7200 &       \\
J1608+0715    &                & 16:08:43.9 & +07:15:09   & 16.60 & 2.88  & 11 Apr 2007  & C1 & HIRESb & 9000  & 11/26 \\
J1712+5755    &                & 17:12:27.74& +57:55:06   & 17.46 & 3.01  & 09 Sep 2004  & C1 & HIRESb & 3600  & X/12 \\
              &                &            &             &       &       & 02 May 2005  & C5 & HIRESb & 3900  & \\
              &                &            &             &       &       & 19 Aug 2006  & C1 & HIRESb & 3900  & \\
              &                &            &             &       &       & 20 Aug 2006  & C1 & HIRESb & 3900  & \\
J1733+5400    &                & 17:33:52.23& +54:00:30   & 17.35 & 3.43  & 02 May 2005  & C5 & HIRESb & 5400  & X/30 \\
              &                &            &             &       &       & 22 Aug 2007  & C1 & HIRESr & 5400  & 35/35\\
J2123-0050    &                & 21:23:29.46& --00:50:53  & 16.43 & 2.26  & 20 Aug 2006  & E3 & HIRESb & 21600 & 30/67 \\
Q2126-1538    &                & 21:29:12.2 & --15:38:41  & 17.3  & 3.27  & 08 Sep 2004  & C1 & HIRESb & 7200  & 10/16 \\
LB2203-1833   &                & 22:06:39.6 & --18:18:46  & 18.4  & 2.73  & 09 Sep 2004  & C1 & HIRESb & 5400  & \\
Q2231$-$00    &LBQS 2231$-$0015& 22:34:08.8 & +00:00:02   & 17.4  & 3.025 & 01 Nov 1995  & C5 & HIRESO & 14400 & 30 \\
SDSS2303-0939 &                & 23:03:01.5 & --09:39:31  & 17.68 & 3.455 & 08 Nov 2005  & C5 & HIRESr & 7200  & 25/29 \\
SDSS2315+1456 &                & 23:15:43.6 & +14:56:06   & 18.52 & 3.377 & 04 Jun 2006  & C5 & HIRESr & 4800  & 16/11 \\
              &                &            &             &       &       & 08 Nov 2005  & C5 & HIRESr & 4400  &  \\
SDSSJ2334-0908&                & 23:34:46.4 & --09:08:12  & 18.03 & 3.317 & 18 Sep 2007  & C1 & HIRESr & 14400 & 28/X \\
Q2355+0108    &                & 23:58:08.6 &  +01:25:06  & 17.5  & 3.40  & 28 Oct 2005  & C5 & HIRESb & 7200  & 21/30 \\
              &                &            &             &       &       & 04 Jan 2006  & C5 & HIRESb & 6300  &  \\
\enddata
\tablenotetext{a}{Magnitude from the SDSS database ($r$-band) or as
  listed in the SIMBAD Astronomical Database ($V$-band).}
\tablenotetext{b}{Decker employed.}
\tablenotetext{c}{Median signal-to-noise per 3.0\kms\ pixel in the
  quasar continuum at $\approx 5000$\AA\ for the old HIRES detector (HIRESO), 
  $\approx 3400/4000$\AA\ for HIRESb, and $\approx 6000/8000$\AA\ for HIRESr.
  An ``X'' indicates no wavelength coverage or that the S/N was compromised by an LLS.}
\end{deluxetable*}

\begin{deluxetable*}{lcccccccccc}
\tablewidth{0pc}
\tablecaption{JOURNAL OF MIKE OBSERVATIONS\label{tab:mike}}
\tabletypesize{\scriptsize}
\tablehead{
\colhead{QSO} & \colhead{Alt. Name} & \colhead{RA} & \colhead{DEC} &
\colhead{$r/V^a$} & \colhead{$z_{em}$} & \colhead{Date} 
& \colhead{Slit$^b$} & \colhead{Exp} & \colhead{S/N$_{\rm blue}^c$} 
& \colhead{S/N$_{\rm red}^d$} \\
&& (J2000) & (J2000) & (mag) &     & (UT)  & ($''$) & (s) & (pix$^{-1}$) & (pix$^{-1}$)}
\startdata
Q0001-2340    &                & 00:03:45.0 & --23:23:46  & 16.7      & 2.262 & 10 Sep 2005 & 1.0  & 3000 &  3/27 & 16/19 \\
J0103--3009   & LBQS0101--3025 & 01:03:55.3 & --30:09:46  & 17.6  & 3.15  & 02 Sep 2004 & 1.0  & 2400 & 7/9   & 8/10 \\
              &                &            &             &       &       & 04 Sep 2004 & 1.0  & 2400 & 4/7   & 7/8  \\
SDSSJ0106+0048 &                & 01:06:19.2 & +00:48:23.3 & 19.03 & 4.449 & 26 Aug 2003 & 1.0  & 8000 & X/X & 8/9\\
SDSSJ0124+0044 &                & 01:24:03.8 & +00:44:32.7 & 17.9  & 3.834 & 28 Aug 2003 & 1.0  & 8000 & X/4 & 24/17\\
SDSSJ0209-0005&                 & 02:09:50.7 & --00:05:06  & 16.9      & 2.856 & 10 Sep 2005 & 1.0  & 5700 & X/10 & 11/9\\
SDSSJ0244-0816   &                 & 02:44:47.8 & --08:16:06  & 18.2 & 4.068 & 26 Aug 2003 & 1.0  & 5500 & X/2 & 24/12 \\
HE0340-2612  &                 & 03:42:27.8 &  --26:02:43 & 17.4  & 3.14  & 02 Sep 2004 & 1.0  & 2400 & 6/11  & 12/19 \\
                     &                 &            &            &       &       & 04 Sep 2004 & 1.0  & 2400  \\
SDSSJ0344-0653   &                 & 03:44:02.8 & --06:53:00 & 18.64 & 3.957 & 28 Aug 2003 & 1.0  & 3000 & X/X & 12/8\\
SDSS0912+0547   &                 & 09:12:10.35& +05:47:42  & 18.05   & 3.248    & 10 May 2004 & 0.7  &  3600 & 2/3   & 6/X \\
SDSSJ0942+0422   &                 & 09:42:02.0 & +04:22:44   & 17.18 & 3.28  & 03 Apr 2003 & 0.7   & 6000 &    X/9   & 14/XX \\ 
HE0940-1050  &                 & 09:42:53.2 & --11:04:22 & 16.6    & 3.08  & 08 May 2004 & 1.0    & 7200 & X/40   & 35/X   \\
SDSSJ0949+0335   &                 & 09:49:32.3 & +03:35:31   & 18.1  &  4.05 & 04 Apr 2003 & 0.7  & 4000 & 4/4   & 12/8 \\ 
             &                 &            &             &       &       & 05 Apr 2003 & 0.7  & 4000 & 3/5   & 11/8 \\
             &                 &            &             &       &       & 06 Apr 2003 & 0.7  & 4000 & 2/5   & 11/10 \\
SDSSJ1025+0452   &                 & 10:25:09.6 & +04:52:46   & 18.0  &  3.24 & 05 Apr 2003 & 0.7  & 4000 & 1/7   & 12/9 \\
             &                 &            &             &       &       & 06 Apr 2003 & 0.7  & 4000 & 1/7   & 12/9 \\
SDSSJ1028--0046  &                 & 10:28:32.1 & --00:46:07  & 17.94 & 2.86 &  12 May 2004 & 1.0  & 6484 & X/8   & 15/6\\
SDSSJ1032+0541   &                 & 10:32:49.9 & +05:41:18.3 & 17.2   & 2.829    & 10 May 2004 & 1.0 & 7200 & 4/13  &  21/X\\
SDSSJ1034+0358   &                 & 10:34:56.3 & +03:58:59   & 17.9  &  3.37 & 04 Apr 2003 & 0.7  &12000 & 1/4   & 15/12 \\
Q1100--264   &                 & 11:03:25.6 & --26:45:06  & 16.02    &  2.145 & 16 May 2005 & 1.0   & 2000 & 4/17  & 16/15 \\
             &                 &            &             &       &       & 18 May 2005 & 1.0  &   2000 & 14/37 & 12/9 \\
HS1104+0452  &                 & 11:07:08.4 & +04:36:18   & 17.48    & 2.66   & 19 May 2005 & 1.0  & 2000 & 2/9   & 12/14 \\
SDSSJ1110+0244   &                 & 11:10:08.6 & +02:44:58   & 18.3  &  4.12 & 05 Apr 2003 & 0.7  & 8000 & 1/3   & 10/7 \\
SDSSJ1136+0050   &                 & 11:36:21.0 & +00:50:21   & 18.1  & 3.43  & 06 Apr 2003 & 0.7  & 8000 & 2/7   & 14/13 \\
SDSSJ1155+0530   &                 & 11:55:38.6 & +05:30:50   & 18.1  & 3.47  & 10 May 2004 & 1.0  & 3600 & 3/9   & 12/10 \\
SDSSJ1201+0116   &                 & 12:01:44.4 & +01:16:11   & 17.5  & 3.23  & 03 Apr 2003 & 0.7  & 8000 & 1/5   & 15/10 \\
LB1213+0922  &                 & 12:15:39.6 & +09:06:08   & 18.26 & 2.723 & 13 May 2004 & 0.7  & 7200   & 3/10      & 10/10  \\
SDSSJ1249--0159  &                 & 12:49:57.2 & --01:59:28  & 17.8  & 3.64  & 06 Apr 2003 & 0.7  & 8000 & 1/13  & 16/15 \\
SDSSJ1307+0422   &                 & 13:07:56.7 & +04:22:15   & 18.0  & 3.02  & 09 May 2004 & 0.7  & 7200 & 2/6   & 10/10 \\
SDSSJ1337+0128   &                 & 13:37:57.9 & +02:18:20   & 18.13 & 3.33  & 12 May 2004 & 1.0 & 6800 & X/10 & 10/9 \\  
SDSSJ1339+0548   &                 & 13:39:42.0 & +05:48:22   & 17.8  & 2.98  & 10 May 2004 & 1.0  & 7200 & 6/13  & 14/12 \\
SDSSJ1402+0146   &                 & 14:02:48.1 & +01:46:34   & 18.8  & 4.16  & 05 Apr 2003 & 0.7   & 8000 & 1/2   & 12/9 \\
SDSSJ1429--0145  & Q1426--0131     & 14:29:03.0 & --01:45:18  & 17.8  & 3.42 & 06 Apr 2003 & 0.7  & 8000 & 2/12  & 13/11 \\
             &                 &            &             &       &       & 17 May 2005 & 1.0   & 8000 &       &       \\
Q1456--1938  &                 & 14:56:50.0 & --19:38:53  & 18.7  & 3.16  & 18 May 2005 & 0.7  &   7200 & 5/10  & 14/22 \\
SDSSJ1503+0419   &                 & 15:03:28.9 &  +04:19:49  & 18.1  & 3.66  & 09 May 2004 & 0.7  & 7200 & 1/3   & 8/7   \\ 
SDSSJ1558--0031  &                 & 15:58:10.2 & --00:31:20  & 17.6  & 2.83  & 06 Apr 2003 & 0.7   & 6000 & 1/6   & 12/8  \\
             &                 &            &             &       &       & 10 May 2004 & 1.0  & 8000 & 3/11  & 18/16 \\
Q1559+0853   &                 & 16:02:22.6 & +08:45:36.5 & 17.3    & 2.269 & 17 May 2005 & 1.0 & 4000 & 4/17 & 11/17 \\
SDSSJ1621-0042   &                 & 16:21:16.9 & --00:42:50  & 17.4  & 3.70  & 03 Apr 2003 & 0.7  & 6000 & X/2   & 11/10 \\   
             &                 &            &             &       &       & 05 Apr 2003 & 0.7   & 3000 & X/7   & 16/12 \\
             &                 &            &             &       &       & 06 Apr 2003 & 0.7  & 3600 & X/7   & 18/14 \\
             &                 &            &             &       &       & 08 May 2004 & 1.0  & 3600 & 3/11  &       \\
PKS2000--330 & Q2000--330      & 20:03:24.1 & --32:51:44  & 17.3  & 3.77  & 02 Sep 2004 & 1.0  & 4800 & 12/35 & 24/19 \\
B2050--359   &                 & 20:53:44.6 & --35:46:52  & 17.7  & 3.49  & 18 May 2005 & 1.0 & 4800 & X/8 & 10/10 \\
Q2126-1538   &                 & 21:29:12.2 & --15:38:41  & 17.3  & 3.27  & 05 Sep 2004 & 1.0  & 4800  &  9/25 & 19/23 \\
HE2215--6206 &                 & 22:18:51.3 & --61:50:54  & 17.5  & 3.32  & 02 Sep 2004 & 1.0  & 2400 &  8/20 & 16/14 \\
             &                 &            &             &       &       & 04 Sep 2004 & 1.0  & 4000 &  7/17 & 17/19 \\
SDSSJ2303-0939&                 & 23:03:01.4 & --09:39:30  & 17.68 & 3.455 & 28 Aug 2003 & 1.0  & 8000 & X/14 & 23/20 \\
HE2314--3405 &                 & 23:16:43.2 & --33:49:12  & 16.9  & 2.96  & 02 Sep 2004 & 1.0  & 2400 &  2/11 & 13/11 \\
SDSSJ2346-0016&                 & 23:46:25.7 & --00:16:00  & 17.77 & 3.49  & 27 Aug 2003 & 1.0  & 8000 & X/14 & 21/26\\
             &                 &            &             &       &       & 28 Aug 2003 & 1.0  & 3000 &  \\
HE2348--1444 &                 & 23:48:55.4 & --14:44:37  & 16.7  & 2.93  & 02 Sep 2004 & 1.0  & 2400 &  14/22& 30/33 \\
HE2355--5457 &                 & 23:58:33.4 & --54:40:42  & 17.1  & 2.94  & 02 Sep 2004 & 1.0  & 2400 &  17/7 & 13/15 \\

		    
\enddata
\tablenotetext{a}{Magnitude from the SDSS database ($r$-band) or as
  listed in the SIMBAD Astronomical Database ($V$-band).}
\tablenotetext{b}{Slit width employed.  For the blue (red) side, a $1''$
  slit yields a FWHM resolution of 10.7 (13.6) \kms\ for a source that fills the
  slit.}
\tablenotetext{c}{Median signal-to-noise per 3.0\kms\ pixel in the
  quasar continuum at $\approx 3400/4000$\AA. An X designates no flux.}
\tablenotetext{d}{Median signal-to-noise per 4.2\kms\ pixel in the
  quasar continuum at $\approx 6000/8000$\AA. An X designates no flux.}
\end{deluxetable*}

\begin{figure*}
\includegraphics[width=7in]{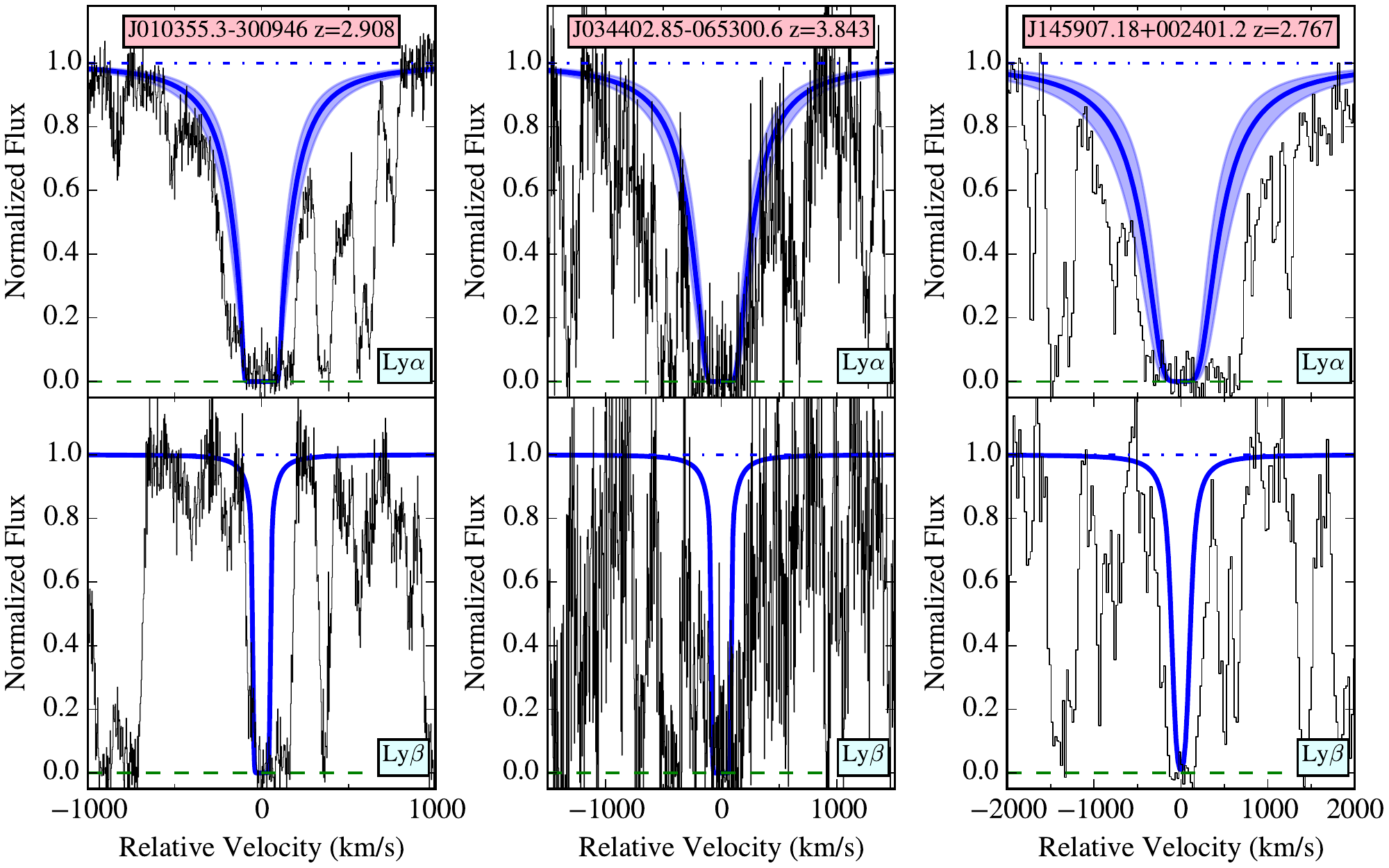}
\caption{\ion{H}{1} \lya\ (top row) and \lyb\ (bottom row)
  profiles for three LLSs with $\mnhi
  \ge 10^{19} \cm{-2}$.  The damping wings of \lya\ are well-resolved
  in these SLLSs and the blue curves indicate for the best-estimates and
  uncertainty of the \nhi\ values.  
  These are well constrained, even in poor S/N data.
}
\label{fig:HI_SLLS}
\end{figure*}

\subsection{Observations}
\label{sec:obs}

We present data obtained at the W.M. Keck and Las Campanas
Observatories using the twin 10\,m Keck~I and Keck~II telescopes and
the twin 6.5\,m Baade and Clay telescopes.  Altogether, we used four
spectrometers:
 (1) the High Resolution Echelle Spectrometer
 \citep[HIRES;][]{vogt94};
 (2) the Echellette Spectrograph and Imager \citep[ESI;][]{ESI};
 (3) the Magellan Inamori Kyocera Echelle \citep[MIKE;][]{MIKE};
and
 (4) the Magellan Echellette Spectrograph \citep[MagE;][]{MAGE}.

The MagE spectra were presented in \cite{fop+13} and we refer
the reader to that manuscript for details on the observations and data
reduction.  Similarly the ESI observations have been published
previously in a series of papers \citep{p03_esi,pwh+07,opb+07,rafelski+12}.

Observing logs for the HIRES and MIKE spectra are
provided in Tables~\ref{tab:hires} and \ref{tab:mike}.
A significant fraction of these data have been analyzed previously
\citep[e.g.][]{opb+07,fpl+08,marcel13}, but not for a comprehensive 
LLS survey.

\subsection{Data Reduction}
\label{sec:redux}

The HIRES spectra were reduced with the
HIRedux\footnote{http://www.ucolick.org/$\sim$xavier/HIRedux/index.html}
software package, primarily as part of the KODIAQ
project \citep{lehner+14}. 
Briefly, each spectral image was bias-subtracted, flat-fielded with
pixel flats, and wavelength-calibrated with corresponding ThAr
frames.  The echelle orders were traced using a traditional flat-field
spectral image.  The sky background was subtracted with a b-spline algorithm
\citep[e.g.][]{mage_pipeline},
and the quasar flux was further traced and optimally
extracted with standard techniques. 
These spectra were flux normalized with a high-order Legendre
polynomial and co-added after weighting by the median S/N
of each order.  This yields an individual, wavelength-calibrated
spectrum for each night of observation in the vacuum and heliocentric
frame.  When possible, we then combined spectra
from quasars observed on multiple nights with the same instrument
configuration.

Processing of the MIKE spectra used the MIRedux
package now bundled within the XIDL software
package\footnote{http://www.ucolick.org/$\sim$xavier/IDL/index.html}.
This pipeline uses algorithms similar to HIRedux.  
The primary difference is that the flux is estimated
together with the sky using a set of b-spline models which is
demanded by the short $5''$ slits employed with MIKE.  
In addition, these data
were fluxed prior to coaddition using the reduced spectrum of a
spectrophotometric standard (taken from the same night in
most cases).  Therefore, we provide both fluxed and normalized
spectra from this instrument.

Details on the data reduction of ESI and MagE spectra are provided in
previous publications \citep{p03_esi,fop+13}.

All of the reduced and calibrated spectra are available on the
project's website\footnote{http://www.ucolick.org/$\sim$xavier/HD-LLS/DR1}.
The Keck/HIRES spectra will also be provided in the first data release
of the KODIAQ project (Lehner et al.\ 2015).

\begin{figure*}
\includegraphics[width=7in]{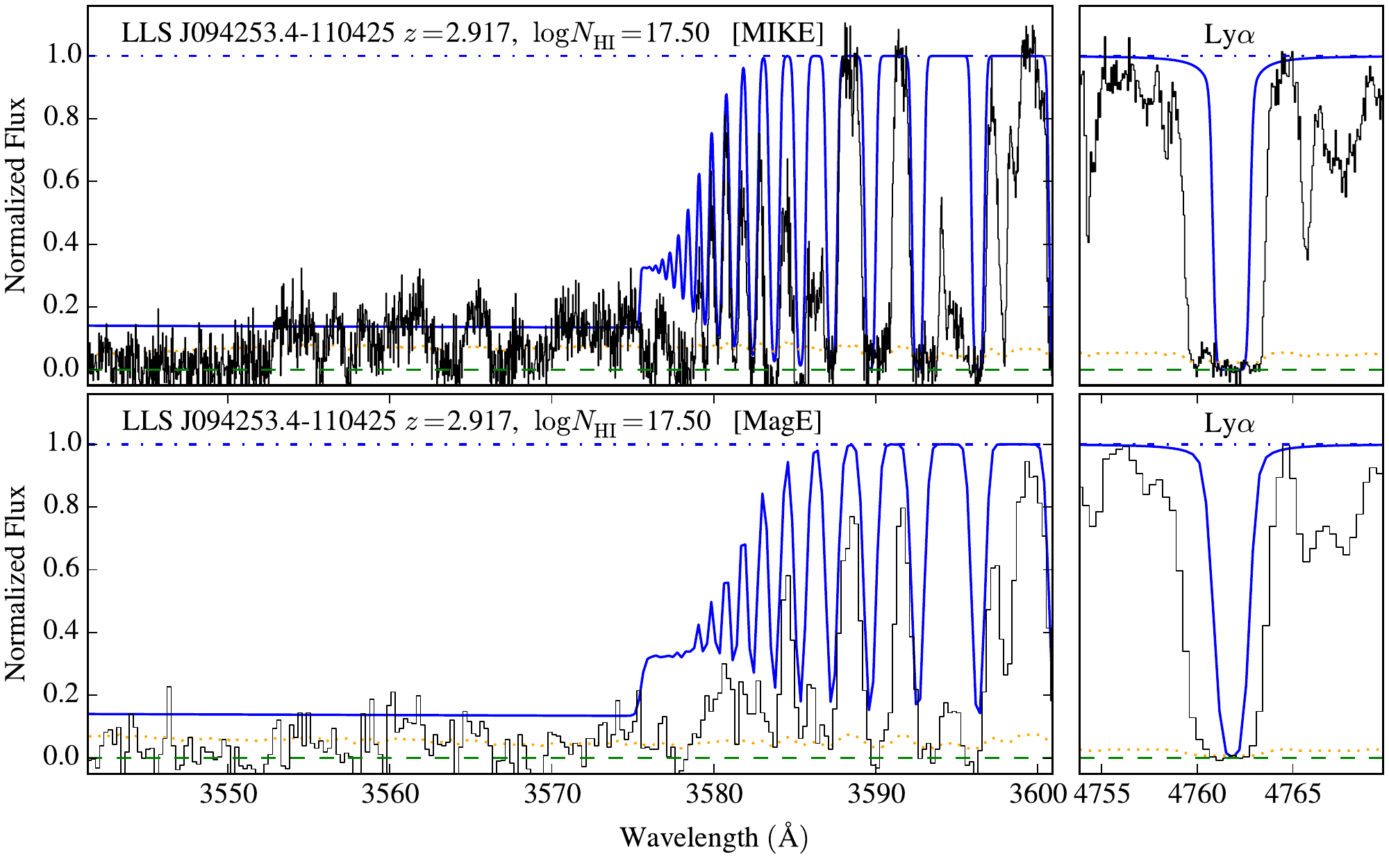}
\caption{Plots of the Lyman limit (left) and the \lya\ (right) profile for
  the LLS at $z=2.917$ toward J0924--1104.  We observed this system
  with both the MIKE (top) and MagE (bottom) spectrometers at LCO.  The
  spectra show a break in the flux at $\lambda \approx 3585$\AA\ but
  residual flux down to $\lambda \approx 3500$\AA\ where the Lyman
  limit from a lower redshift absorber occurs.  By modeling the flux
  decrement of the $z=2.917$ LLS, we establish a precise
  estimate of its total \nhi\ value.  It is also evident that the \ion{H}{1}
  \lya\ profile alone offers very little constraint on \nhi.
}
\label{fig:HI_pLLS}
\end{figure*}

\section{LLS Definition}
\label{sec:define}

Before proceeding to analysis of the sample, we strictly
define the Lyman limit system.  There are three aspects to the
definition:
  \begin{enumerate}
    \item The velocity interval analyzed, which also corresponds to a
      finite redshift window.
    \item The \nhi\ value of the system.
    \item The spatial proximity of the LLS to other astrophysical
      objects (e.g.\ the background quasar or a foreground DLA).
  \end{enumerate}
Of these three, the first has received the least attention by the
community yet may be the most important.  
Establishing a precise
definition, however, is largely arbitrary despite the fact that it
may significantly
impact the studies that follow.  This includes the assessment
of gas kinematics \citep{pw97}, metallicity \citep{ppo+10,fop11}, and even the
contribution of LLSs to the cosmic mean free path \citep{pmo+14}.  In
this paper (and future publications), we adopt an
observationally-motivated velocity interval of $\pm 500 
\mkms$ centered on the peak optical depth \zpeak\ of the \ion{H}{1}
Lyman series. 
Frequently, we estimate \zpeak\ from the peak optical depth of
a low-ion, metal-line transition.
An LLS, then, is all of the optically thick
gas at $|\delta v| < 500 \mkms$ from \zpeak.  
In practice, we have not simply summed the \ion{H}{1} column densities
of all \lya\ absorbers within this interval.  Instead, we have adopted
the integrated \nhi\ estimate from the Lyman limit decrement or adopt
\nhi\ from the analysis of damping in the \lya\ profiles (see the next
section for more detail).
As an example of the latter, we treat the two absorbers at
$z=3.1878$ and $z=3.1917$ towards PKS2000-330 \citep{ppo+10} 
as a single LLS.  Similarly, we sum metal-line absorption identified
within the interval although it rarely is detected in intervals
that exceeds 200\kms.
Moreover, this window was adjusted further to exclude absorption from
unrelated (e.g higher or lower redshift) systems.
While this is an observationally driven definition, we note that
it should also capture even the largest peculiar motions
within dark matter halos at $z \sim 2$.

With $|\delta v| < 500 \mkms$ as the first criterion, we define an LLS
as any combination of systems with $\mnhi \ge 10^{17.3} \cm{-2}$  
within that interval;  this yields an integrated optical
depth at the Lyman limit $\mtll \ge 1$.  
In practice, we distinguish the LLS from DLAs by requiring
that $\mnhi < 10^{20.3} \cm{-2}$.  
Systems with $10^{16} < \mnhi < 10^{17.3} \cm{-2}$
are referred to as partial LLSs or pLLSs, and are excluded from analysis
in this manuscript.
Last, we refer to an LLS within 3000\,\kms\ of the background quasar
as a proximate LLS or PLLS \citep{phh08}.
There are 5 PLLSs within our sample satisfying this definition, all
with velocity separations of at least $2000 \mkms$ from the reported
quasar redshifts.
Altogether, we present measurements for \nhdlls~LLSs at redshifts
$\mzlls = 1.76-4.39$ and with $\mnhi = 10^{17.3} - 10^{20.25} \cm{-2}$.
Here and in future papers we refer to this dataset as the
high-dispersion LLS sample (HD-LLS Sample).  We will
augment this sample in the years to follow via our web site.

\begin{figure*}
\includegraphics[width=7in]{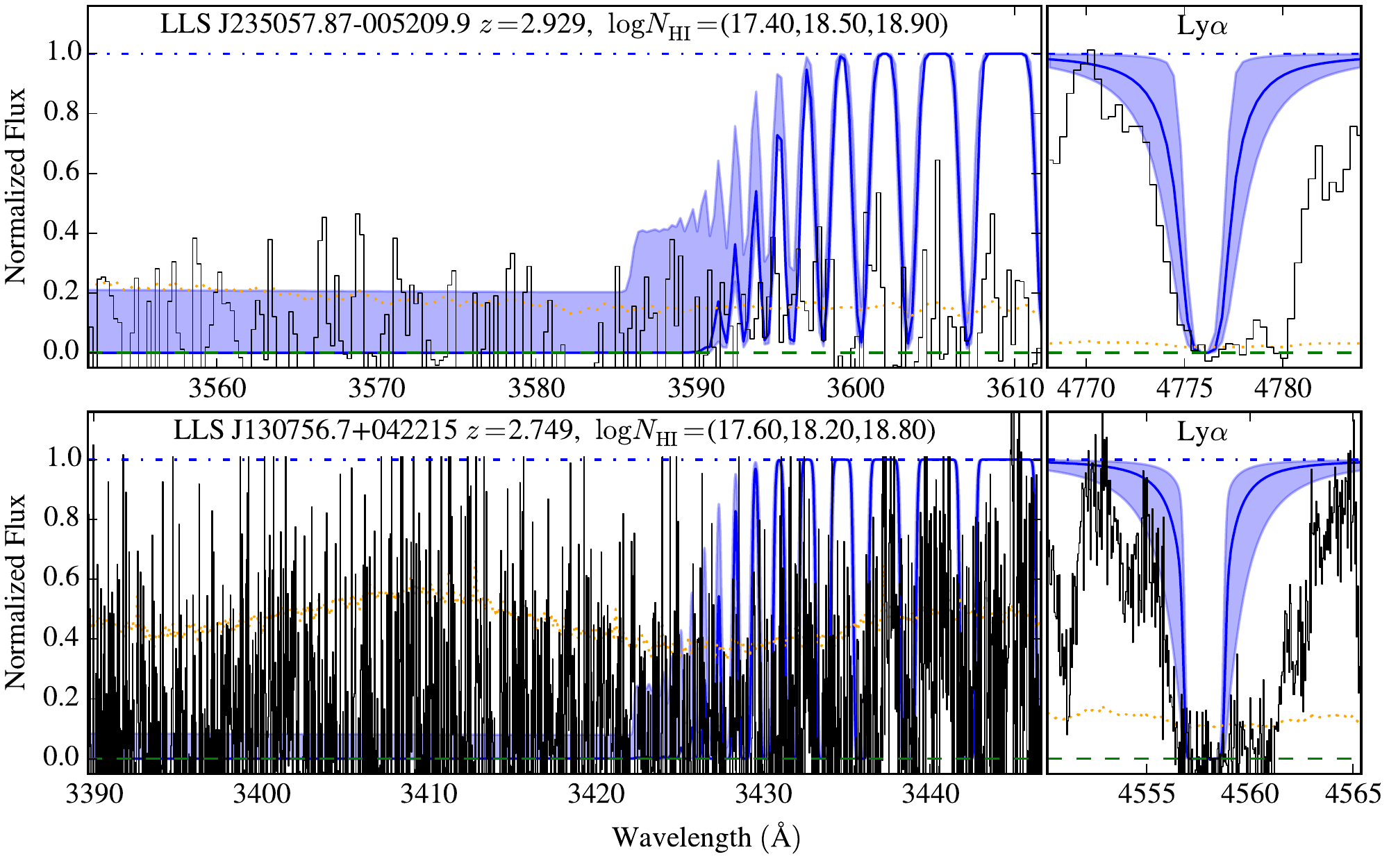}
\caption{Two examples of LLSs with a strong Lyman limit break ($\mtll >
  1$) yet weak or absent damping of \ion{H}{1} \lya.  In these
  examples, we only estimate bounds on the \nhi\ values which
  can span an order-of-magnitude uncertainty.
}
\label{fig:HI_ambig}
\end{figure*}

\section{\nhi\ Analysis}
\label{sec:NHI}

Although the continuum opacity of the Lyman limit generates an
unambiguous signature in a quasar spectrum, it is generally
challenging to precisely estimate the \ion{H}{1}
column density \nhi\ for a given LLS system.  
This follows simply from the fact that $\exp(-\mtll) \ll 1$ for 
$\mnhi > 10^{18} \cm{-2}$ and all of
the \ion{H}{1} Lyman series lines are
on the saturated portion of the curve-of-growth for 
$\mnhi < 10^{19} \cm{-2}$.
Furthermore, 
the damping of \lya\ is difficult to measure
for $\mnhi \ll 10^{20} \cm{-2}$, especially in low
S/N spectra or at $z>3$ where IGM blending is substantial.

Our approach to identifying LLS and estimating their \nhi\ values
involved two relatively distinct procedures.  For LLS
with large \nhi\ values ($>10^{19} \cm{-2}$), we
searched each spectrum for absorption features with large equivalent
widths \ew\ characteristic of a damped \lya\ line (i.e. $\mew \gg
1$\AA).  We then considered whether these
candidates could be related to a broad absorption line (BAL)
system associated to the background quasar or \lyb\ associated to a
higher redshift DLA.  Any such coincidences were eliminated.
For the remaining candidates, we performed an analysis of the \lya\
profile by overplotting a series of Voigt profiles with $\mnhi >
10^{18.5} \cm{-2}$, adjusting the local continuum by-eye as
warranted.  
When low-ion metal absorption was detected near the
approximate centroid of \lya, we centered the model
to its peak optical depth and refined the \nhi\ value
accordingly.  
We did not, however, require the positive detection of 
metal-line absorption.
In all cases, the Doppler parameter of the model \lya\
line was set to 30\,\kms. 

For cases were the S/N was deemed sufficient and line-blending not too
severe, we estimated (by visual inspection) 
a `best' \nhi\ value and corresponding
$1\sigma$ uncertainties.  Although this procedure is rife
with human interaction, 
we maintain that it offers the most robust assessment (to
date) for \nhi\ estimation.  This is because the dominant
uncertainties are systematic (e.g.\ continuum placement and
line-blending), which are difficult to estimate statistically.
Figure~\ref{fig:HI_SLLS} shows three examples of LLSs with damped
\lya\ lines giving precisely estimated \nhi\ values.
Such systems are commonly referred to as super LLSs (SLLSs) or sub-DLAs.

\begin{deluxetable}{lccccccccc}
\tablewidth{0pc}
\tablecaption{\nhi\ ESTIMATES FOR THE HD-LLS SAMPLE\label{tab:HI}}
\tabletypesize{\footnotesize}
\tablehead{\colhead{Quasar} & \colhead{$z_{\rm peak}^a$} & \colhead{$N_{\rm HI}^{\rm low}$}
& \colhead{$N_{\rm HI}^{\rm high}$}
& \colhead{$N_{\rm HI}^{\rm adopt}$}
& \colhead{$flg_{\rm HI}^b$}}
\startdata
J1608+0715&1.7626&19.10&19.70&$19.40_{-0.30}^{+0.30}$&1\\
J0953+5230&1.7678&20.00&20.20&$20.10_{-0.10}^{+0.10}$&1\\
J0927+5621&1.7749&18.90&19.10&$19.00_{-0.10}^{+0.10}$&1\\
J1509+1113&1.8210&18.00&19.00&$18.50_{-0.50}^{+0.50}$&2\\
J101939.15+524627&1.8339&18.80&19.40&$19.10_{-0.30}^{+0.30}$&1\\
Q1100-264&1.8389&19.25&19.55&$19.40_{-0.15}^{+0.15}$&1\\
J1159-0032&1.9044&19.90&20.20&$20.05_{-0.15}^{+0.15}$&1\\
Q0201+36&1.9548&19.90&20.30&$20.10_{-0.20}^{+0.20}$&1\\
J0828+0858&2.0438&19.80&20.00&$19.90_{-0.10}^{+0.10}$&1\\
J2123-0050&2.0593&19.10&19.40&$19.25_{-0.15}^{+0.15}$&1\\
Q1456-1938&2.1701&19.55&19.95&$19.75_{-0.20}^{+0.20}$&1\\
J034024.57-051909&2.1736&19.15&19.55&$19.35_{-0.20}^{+0.20}$&1\\
Q0001-2340&2.1871&19.50&19.80&$19.65_{-0.15}^{+0.15}$&1\\
SDSS1307+0422&2.2499&19.85&20.15&$20.00_{-0.15}^{+0.15}$&1\\
J1712+5755&2.3148&20.05&20.35&$20.20_{-0.15}^{+0.15}$&1\\
Q2053-3546&2.3320&18.75&19.25&$19.00_{-0.25}^{+0.25}$&1\\
Q2053-3546&2.3502&19.35&19.85&$19.60_{-0.25}^{+0.25}$&1\\
Q1456-1938&2.3512&19.35&19.75&$19.55_{-0.20}^{+0.20}$&1\\
J1131+6044&2.3620&19.90&20.20&$20.05_{-0.15}^{+0.15}$&1\\
Q1206+1155&2.3630&20.05&20.45&$20.25_{-0.20}^{+0.20}$&1\\
HE2314-3405&2.3860&18.80&19.20&$19.00_{-0.20}^{+0.20}$&1\\
Q0301-005&2.4432&19.75&20.05&$19.90_{-0.15}^{+0.15}$&1\\
HS1345+2832&2.4477&19.70&20.00&$19.85_{-0.15}^{+0.15}$&1\\
J1035+5440&2.4570&19.40&19.90&$19.65_{-0.25}^{+0.25}$&1\\
Q1337+11&2.5080&20.00&20.30&$20.15_{-0.15}^{+0.15}$&1\\
SDSS0912+0547&2.5220&19.15&19.55&$19.35_{-0.20}^{+0.20}$&1\\
SDSS0209-0005&2.5228&18.90&19.20&$19.05_{-0.15}^{+0.15}$&1\\
LB1213+0922&2.5230&20.00&20.40&$20.20_{-0.20}^{+0.20}$&1\\
Q0207-003&2.5231&18.80&19.20&$19.00_{-0.20}^{+0.20}$&1\\
Q0207-003&2.5466&17.60&18.60&$18.10_{-0.50}^{+0.50}$&2\\
HS1104+0452&2.6014&19.70&20.10&$19.90_{-0.20}^{+0.20}$&1\\
J2234+0057&2.6040&19.25&19.75&$19.50_{-0.25}^{+0.25}$&1\\
J115659.59+551308.1&2.6159&18.80&19.30&$19.10_{-0.30}^{+0.30}$&1\\
SDSS1558-0031&2.6300&19.40&19.75&$19.60_{-0.20}^{+0.20}$&1\\
SDSS0157-0106&2.6313&19.25&19.65&$19.45_{-0.20}^{+0.20}$&1\\
Q2126-158&2.6380&19.10&19.40&$19.25_{-0.15}^{+0.15}$&1\\
Q1455+123&2.6481&17.30&19.40&$18.35_{-1.05}^{+1.05}$&2\\
LBQS2231-0015&2.6520&18.70&19.30&$19.10_{-0.40}^{+0.20}$&1\\
SDSS0121+1448&2.6623&19.05&19.40&$19.25_{-0.20}^{+0.15}$&1\\
SDSSJ0915+0549&2.6631&17.50&18.90&$18.20_{-0.70}^{+0.70}$&2\\
SDSSJ2319-1040&2.6750&19.30&19.60&$19.45_{-0.15}^{+0.15}$&1\\
Q0201+36&2.6900&17.50&18.80&$18.50_{-1.00}^{+0.30}$&2\\
LB2203-1833&2.6981&19.85&20.15&$20.00_{-0.15}^{+0.15}$&1\\
SDSSJ1551+0908&2.7000&17.30&17.70&$17.50_{-0.20}^{+0.20}$&1\\
HS1200+1539&2.7080&17.60&18.90&$18.30_{-0.70}^{+0.70}$&2\\
Q1508+087&2.7219&19.00&19.80&$19.40_{-0.40}^{+0.40}$&1\\
PMNJ1837-5848&2.7289&17.50&18.70&$18.10_{-0.60}^{+0.60}$&2\\

\enddata
\tablecomments{All column densities are log$_{10}$. The flag in the final column indicates
the quality of the measurement.  A $flg_{\rm HI} = 1$ corresponds to a more precisely measured value and 
one may assume a Gaussian PDF with the errors reported taken as $1\sigma$ uncertainties.
A $flg_{\rm HI} = 2$ corresponds to a less precisely measured value, and we recommend one adopt a uniform
prior for \nhi\ within the error interval reported.  See text for further details.}
\tablenotetext{a}{Redshift estimates for the peak \ion{H}{1} opacity from metals and Lyman series absorption.}
\end{deluxetable}

We provide the adopted \nhi\ values and error estimates of our SLLS
sample  (\nslls\ systems with $\mnhi \ge 10^{19} \cm{-2}$)
in Table~\ref{tab:HI}.  These represent roughly 2/3 of the total HD-LLS
Sample.  This high fraction occurs because we only require coverage of
\ion{H}{1} \lya\ to identify and analyze these SLLS.
This implies a much larger survey path-length than for the
lower \nhi\ LLS.  In addition, one may identify and analyze multiple
SLLSs 
along a given sightline whereas one is restricted to a single LLS when
the Lyman Limit is central to the analysis.
Because these SLLSs tend to span nearly the entire $\pm 500 \mkms$
window that defines an LLS, it is possible that there is additional,
optically thick gas not included in our \nhi\ estimate.  This will be
rare, however, and the underestimate of \nhi\ should generally be 
much less than 10\%.

\begin{figure}
\includegraphics[width=3.5in]{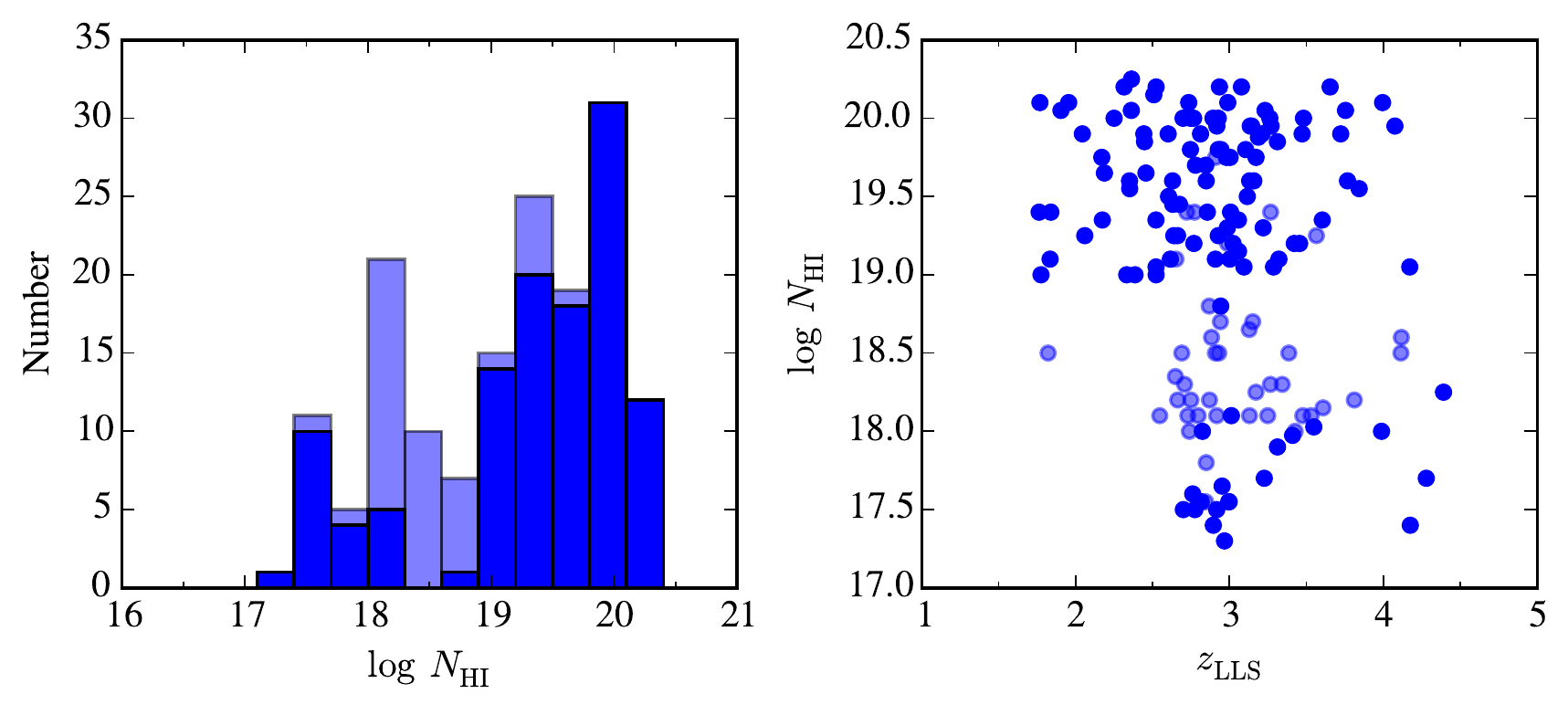}
\caption{(left) Histogram of \nhi\ values for the HD-LLS Sample.
  The ligher bars indicate \nhi\ values with large uncertainty
  ($1\sigma > 0.4$\,dex).
  Nearly two-thirds of the sample have $\mnhi \ge 10^{19} \cm{-2}$ 
  which is a consequence of the
  spectral coverage required to estimate \nhi\ for an LLS (see text).
  (right) Scatter plot of \nhi\ vs.\ $z_{\rm LLS}$ for the
  sample. Again,
  the lighter points indicate LLSs with poorer constraints on the \nhi\
  values (flg=2 in Table~\ref{tab:HI}).  
  Systems with $\mnhi< 10^{19} \cm{-2}$ all have $z_{\rm LLS} > 2.6$
  as coverage of the Lyman limit is required for the \ion{H}{1}
  analysis.
}
\label{fig:NHI_summ}
\end{figure}

In parallel with the search for LLSs having strong \lya\ lines, we
inspected each spectrum for a Lyman limit break.
For those the LLSs that exhibit non-negligible flux at the Lyman limit,
i.e. $\mtll < 3$, a precise \nhi\ estimation may be
recovered.  In practice, such analysis is hampered by poor sky
subtraction and associated IGM absorption that stochastically reduces
the quasar flux through the spectral region near the Lyman limit and
affects continuum placement.
In the following, we have been conservative regarding the systems with
\nhi\ measurements from the Lyman limit flux decrement.
We are currently acquiring additional,
low-dispersion spectra to confirm the flux at the Lyman limit for a
set of the HD-LLS Sample.
Figure~\ref{fig:HI_pLLS} shows one example of a pLLS observed with
both the MIKE and MagE spectrometers.  The flux decrement is obvious
and one also appreciates the value of higher spectral resolution (with
high S/N) for resolving IGM absorption.

For the remainder of the systems identified on the basis of a
Lyman limit break, we adopt conservative bounds (i.e.\
upper and lower limits) to the \nhi\ values.  These are based
primarily on analysis of the \lya\ line and the flux at the Lyman
limit.  The absence of strong damping in the former provides a strict
upper limit to \nhi\ while the latter sets a firm lower limit.
These bounds are provided in Table~\ref{tab:HI}, and
Figure~\ref{fig:HI_ambig} shows two examples of these `ambiguous'
cases. 
In practice, the bounds are often an order-of-magnitude apart, e.g.
$10^{17.7} \cm{-2} < \mnhi < 10^{18.9} \cm{-2}$.  Furthermore, 
it is difficult to estimate the probability distribution function
(PDF) of
\nhi\ within these bounds.  One should not, for example, assume a
Gaussian PDF centered within the bounds with a dispersion of half
the interval.  In fact, we expect that the PDF is much closer to
uniform, i.e.\ equal probability for any \nhi\ value within the
bounds.  This expectation is motivated by current
estimations of the \nhi\ frequency distribution \fnhi\ which argue for
a uniform distribution of 
\nhi\ values for randomly selected systems with
$\mnhi \approx 10^{18} \cm{-2}$ \citep{pow10,omeara13}.
Going forward, we advocate adopting a uniform PDF.

As a cross-check on the analysis, 50 of the sightlines were
re-analyzed by a second author to identify LLSs and estimate their
\nhi\ values.  
With two exceptions, the values between the two
authors agree within the estimated uncertainty 
and we identify no
obvious systematic bias\footnote{In fact the
  formal reduced $\chi^2$ for the comparison is significantly less than
  unity, but this is because the estimated uncertainties include
  systematic error and because each author analyzed the same
  data.}.
These two exceptions have $>0.5$\,dex difference due to differing
definitions used by the two authors and we have adopted the values
corresponding to the strict definition provided in
$\S$~\ref{sec:define}. 
This exercise confirms that the uncertainties are dominated by
systematic effects, not S/N nor the analysis procedures.

\begin{figure*}
\includegraphics[width=7in]{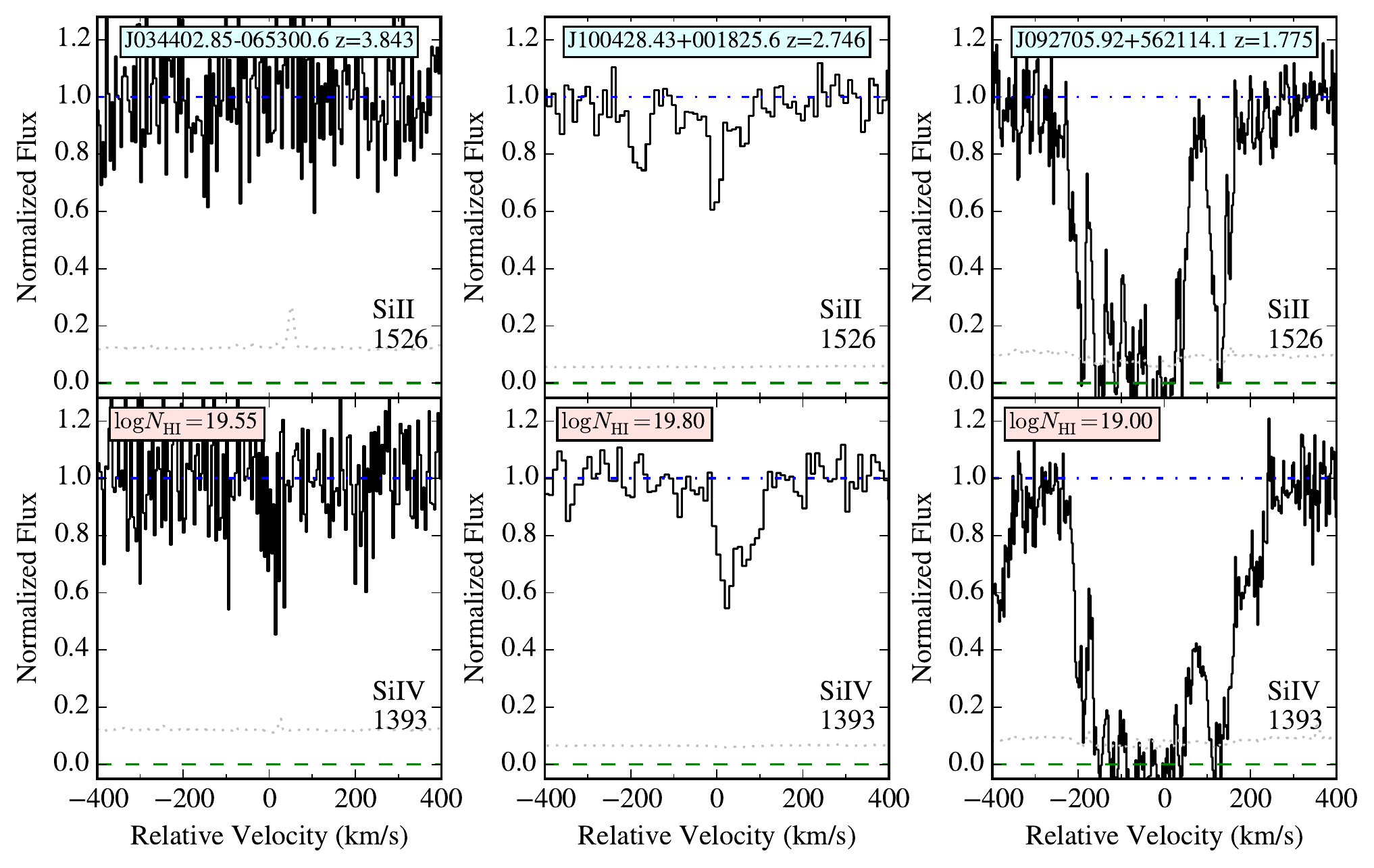}
\caption{\ion{Si}{2} and \ion{Si}{4} transitions for three SLLS
  representative of the full HD-LLS Sample.  Note the large diversity
  in metal-line strength despite the comparable \nhi\ values.  
  These systems have a tendency to show both low and high-ion
  absorption indicative of partially ionized gas.
  The gray dotted line in each panel indicates the estimated $1\sigma$
  error array.
}
\label{fig:ionic_SLLS}
\end{figure*}

\begin{deluxetable*}{lccccccccccccc}
\tablewidth{0pc}
\tablecaption{IONIC COLUMN SUMMARY FOR Si AND C\label{tab:SiCions}}
\tabletypesize{\scriptsize}
\tablehead{\colhead{Quasar} & \colhead{$z_{\rm abs}$} 
& \colhead{\nhi}
& \colhead{$\N{C^{+}}$} & \colhead{$\sigma(\N{C^{+}})$}
& \colhead{$\N{C^{3+}}$} & \colhead{$\sigma(\N{C^{3+}})$}
& \colhead{$\N{Si^{+}}$} & \colhead{$\sigma(\N{Si^{+}})$}
& \colhead{$\N{Si^{3+}}$} & \colhead{$\sigma(\N{Si^{3+}})$}
}
\startdata
J1608+0715&1.7626&$19.40_{-0.30}^{+0.30}$&&&&&15.80&-9.99&&\\
J0953+5230&1.7678&$20.10_{-0.10}^{+0.10}$&15.44&+9.99&15.21&+9.99&15.67& 0.01&14.57&+9.99\\
J0927+5621&1.7749&$19.00_{-0.10}^{+0.10}$&15.40&+9.99&15.40&+9.99&15.58& 0.02&14.84&+9.99\\
J1509+1113&1.8210&$18.50_{-0.50}^{+0.50}$&&&14.83&+9.99&14.21& 0.04&14.17&+9.99\\
J101939.15+524627&1.8339&$19.10_{-0.30}^{+0.30}$&&&14.93&+9.99&15.32& 0.03&14.14&+9.99\\
Q1100-264&1.8389&$19.40_{-0.15}^{+0.15}$&&&14.24& 0.00&13.96& 0.01&13.83& 0.00\\
J1159-0032&1.9044&$20.05_{-0.15}^{+0.15}$&15.38&+9.99&15.22&+9.99&15.14& 0.10&14.54&+9.99\\
Q0201+36&1.9548&$20.10_{-0.20}^{+0.20}$&&&&&15.11& 0.09&&\\
J0828+0858&2.0438&$19.90_{-0.10}^{+0.10}$&15.14&+9.99&14.89&+9.99&15.25& 0.10&14.44&+9.99\\
J2123-0050&2.0593&$19.25_{-0.15}^{+0.15}$&15.11&+9.99&14.60&+9.99&14.60& 0.04&13.96& 0.00\\
Q1456-1938&2.1701&$19.75_{-0.20}^{+0.20}$&&&&&14.84&-9.99&&\\
J034024.57-051909&2.1736&$19.35_{-0.20}^{+0.20}$&14.40&+9.99&13.86& 0.02&13.84& 0.02&13.39& 0.02\\
Q0001-2340&2.1871&$19.65_{-0.15}^{+0.15}$&14.45&+9.99&14.26& 0.01&13.75& 0.03&13.74& 0.01\\
SDSS1307+0422&2.2499&$20.00_{-0.15}^{+0.15}$&&&14.22& 0.03&14.25&+9.99&&\\
J1712+5755&2.3148&$20.20_{-0.15}^{+0.15}$&&&13.36& 0.04&14.08& 0.01&&\\
\enddata
\tablecomments{All column densities are log$_{10}$. When the reported $\sigma=+9.99$, 
the measured value should be taken as a lower limit.  Similarly, $\sigma=-9.99$ indicates that the reported value
refers to an upper limit at 95\%\ c.l.}
\tablecomments{[The complete version of this table is in the electronic edition of the Journal.  The printed edition contains only a sample.]}
\end{deluxetable*}

\begin{figure*}
\includegraphics[width=7in]{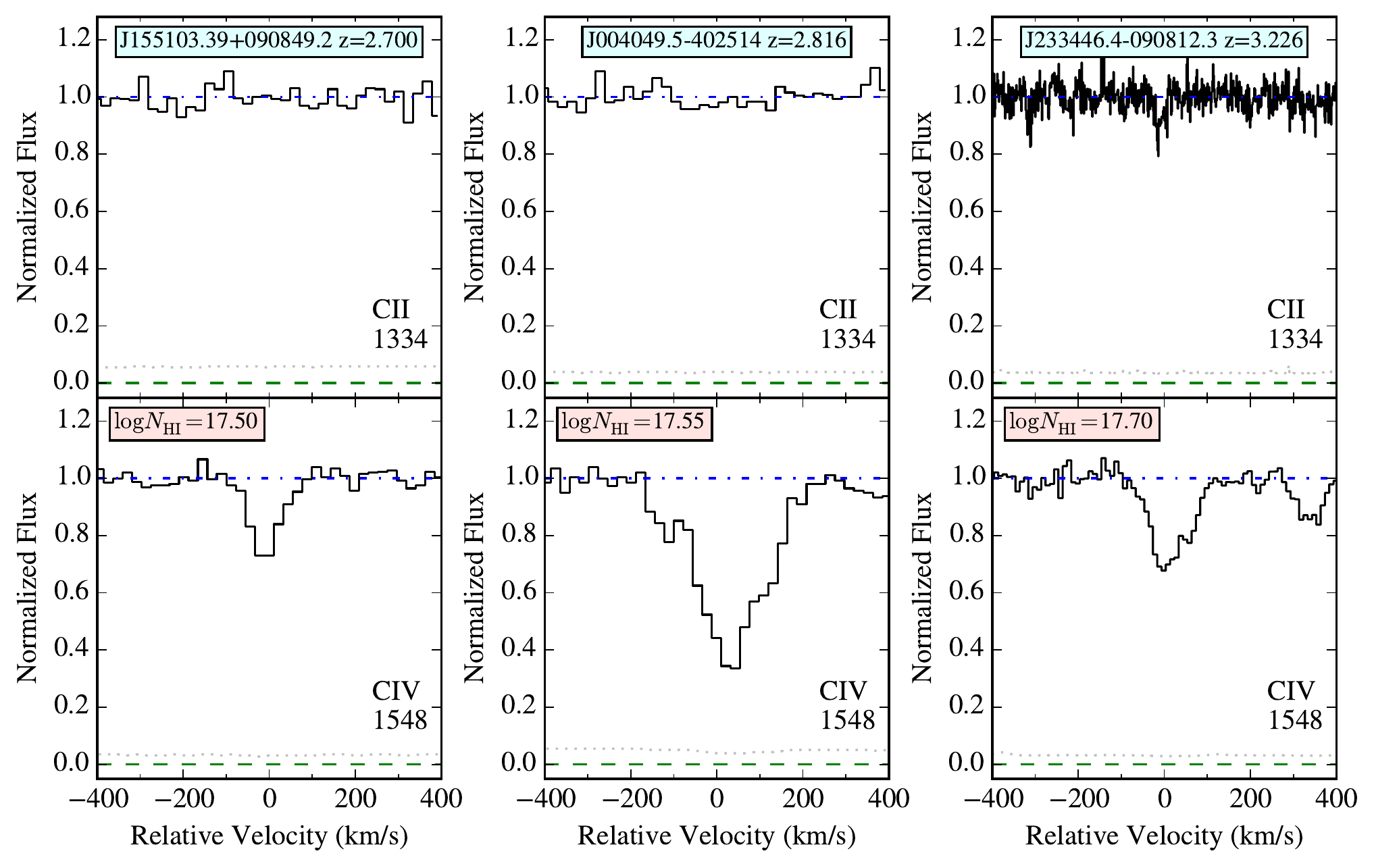}
\caption{\ion{C}{2} and \ion{C}{4} transitions for LLSs with low \nhi\
  values.   Unlike the SLLSs from Figure~\ref{fig:ionic_SLLS}, these
  LLSs have metal absorption that is dominated by high-ions.  There only
  a few cases of LLSs with $\mnhi < 10^{18} \cm{-2}$ and a positive low-ion
  detection. 
}
\label{fig:ionic_pLLS}
\end{figure*}

\begin{deluxetable*}{lccccccccccccc}
\tablewidth{0pc}
\tablecaption{IONIC COLUMN SUMMARY FOR Al, Fe, and O\label{tab:other_ions}}
\tabletypesize{\scriptsize}
\tablehead{\colhead{Quasar} & \colhead{$z_{\rm abs}$} 
& \colhead{\nhi}
& \colhead{$\N{O^{0}}$} & \colhead{$\sigma(\N{O^{0}})$}
& \colhead{$\N{Al^{+}}$} & \colhead{$\sigma(\N{Al^{+}})$}
& \colhead{$\N{Al^{++}}$} & \colhead{$\sigma(\N{Al^{++}})$}
& \colhead{$\N{Fe^{+}}$} & \colhead{$\sigma(\N{Fe^{+}})$}
}
\startdata
J1608+0715&1.7626&$19.40_{-0.30}^{+0.30}$&&&&&13.53& 0.00&&\\
J0953+5230&1.7678&$20.10_{-0.10}^{+0.10}$&15.68&+9.99&13.96&+9.99&13.86& 0.01&14.99& 0.10\\
J0927+5621&1.7749&$19.00_{-0.10}^{+0.10}$&15.63&+9.99&13.92&+9.99&14.05& 0.01&15.28& 0.13\\
J1509+1113&1.8210&$18.50_{-0.50}^{+0.50}$&&&13.12&+9.99&13.04& 0.05&13.76& 0.11\\
J101939.15+524627&1.8339&$19.10_{-0.30}^{+0.30}$&&&13.34&+9.99&13.62& 0.02&14.19& 0.02\\
Q1100-264&1.8389&$19.40_{-0.15}^{+0.15}$&&&12.79& 0.01&12.31& 0.10&13.42& 0.01\\
J1159-0032&1.9044&$20.05_{-0.15}^{+0.15}$&15.66&+9.99&13.98&+9.99&13.82& 0.01&&\\
Q0201+36&1.9548&$20.10_{-0.20}^{+0.20}$&&&13.77&+9.99&13.61& 0.01&&\\
J0828+0858&2.0438&$19.90_{-0.10}^{+0.10}$&15.49&+9.99&&&13.59& 0.02&14.89& 0.04\\
J2123-0050&2.0593&$19.25_{-0.15}^{+0.15}$&&&13.44&+9.99&13.15& 0.01&14.39& 0.00\\
Q1456-1938&2.1701&$19.75_{-0.20}^{+0.20}$&&&13.38&+9.99&12.99& 0.04&14.26& 0.01\\
J034024.57-051909&2.1736&$19.35_{-0.20}^{+0.20}$&14.56&+9.99&12.65& 0.03&12.49& 0.14&&\\
Q0001-2340&2.1871&$19.65_{-0.15}^{+0.15}$&14.16& 0.04&13.00&+9.99&12.40&-9.99&13.11& 0.03\\
SDSS1307+0422&2.2499&$20.00_{-0.15}^{+0.15}$&&&13.04&+9.99&12.80& 0.07&14.18& 0.04\\
J1712+5755&2.3148&$20.20_{-0.15}^{+0.15}$&&&12.56& 0.02&12.39& 0.06&13.65& 0.03\\
\enddata
\tablecomments{All column densities are log$_{10}$. When the reported $\sigma=+9.99$, 
the measured value should be taken as a lower limit.  Similarly, $\sigma=-9.99$ indicates that the reported value
refers to an upper limit at 95\%\ c.l.}
\tablecomments{[The complete version of this table is in the electronic edition of the Journal.  The printed edition contains only a sample.]}
\end{deluxetable*}

Table~\ref{tab:HI} lists the adopted \nhi\ value, errors on this value, the
bounds on \nhi, and a flag indicating whether one would assume a
normal or uniform PDF. 
Figure~\ref{fig:NHI_summ} shows a histogram of the adopted \nhi\
values and a scatter plot against \zlls.
It is evident that the HD-LLS Sample is weighted towards higher \nhi\
values and $z \sim 3$.

\section{Ionic Column Densities}
\label{sec:ionic}

For each of the HD-LLS Sample, we inspected the spectra for associated
metal-line absorption.  Emphasis was placed on transitions with
observed wavelengths redward of the \lya\ forest.
A velocity interval was estimated for the column density measurements
based on the cohort of transitions detected.  
Velocity plots were generated and inspected to search for
line-blending.  Severely blended lines were eliminated from analysis and
intermediate/weak cases were measured but reported as upper limits.
All of these assignments were vetted by JXP, MF, and JMO.
Figures~\ref{fig:ionic_SLLS} and \ref{fig:ionic_pLLS} show the
\ion{Si}{2}~1526/\ion{Si}{4}~1393 transitions for three representative
SLLSs and the
\ion{C}{2}~1334/\ion{C}{4}~1548 transitions for three LLSs
with $\mnhi \lesssim 10^{17.7} \cm{-2}$.
These data indicate a great diversity of line-strengths for these
transitions within the SLLS sample. We also conclude that
metal-absorption is dominated by high-ions in 
the lower \nhi\ systems. 

Column densities were measured using the apparent optical depth method
\citep[AODM;][]{savage91} which gives accurate results for unsaturated
line profiles.  On the latter point, the echelle data (MIKE,HIRES)
have sufficiently high resolution to directly assess line-saturation,
i.e.\ only profiles with minimum normalized
flux \fmin\ less than 0.1 may be saturated. 
For the echellette data (MagE, ESI), however, line-saturation is a
concern \citep[e.g.][]{p03_esi}.
In general, we have proceeded conservatively by treating most lines as
saturated when $\mfmin < 0.5$.   For many of the ions analyzed in
these LLSs, we observe multiple transitions with differing
oscillator strengths and have further assessed
line-saturation from the cohort of measurements.

Uncertainties were estimated from standard propagation of error, which
does not include error from continuum placement.  To be conservative,
we adopt a minimum uncertainty of 0.05\,dex to the measurements from a
given transition.
When multiple transitions from the same ion were measured (e.g.\
\ion{Si}{2}~1304 and \ion{Si}{2}~1526 for Si$^+$),
we calculate the ionic column density from the weighted mean.
Otherwise, we adopt the measurement from the single 
transition or a limit from the cohort emphasizing positive
detections.

\begin{figure}
\includegraphics[width=3.5in]{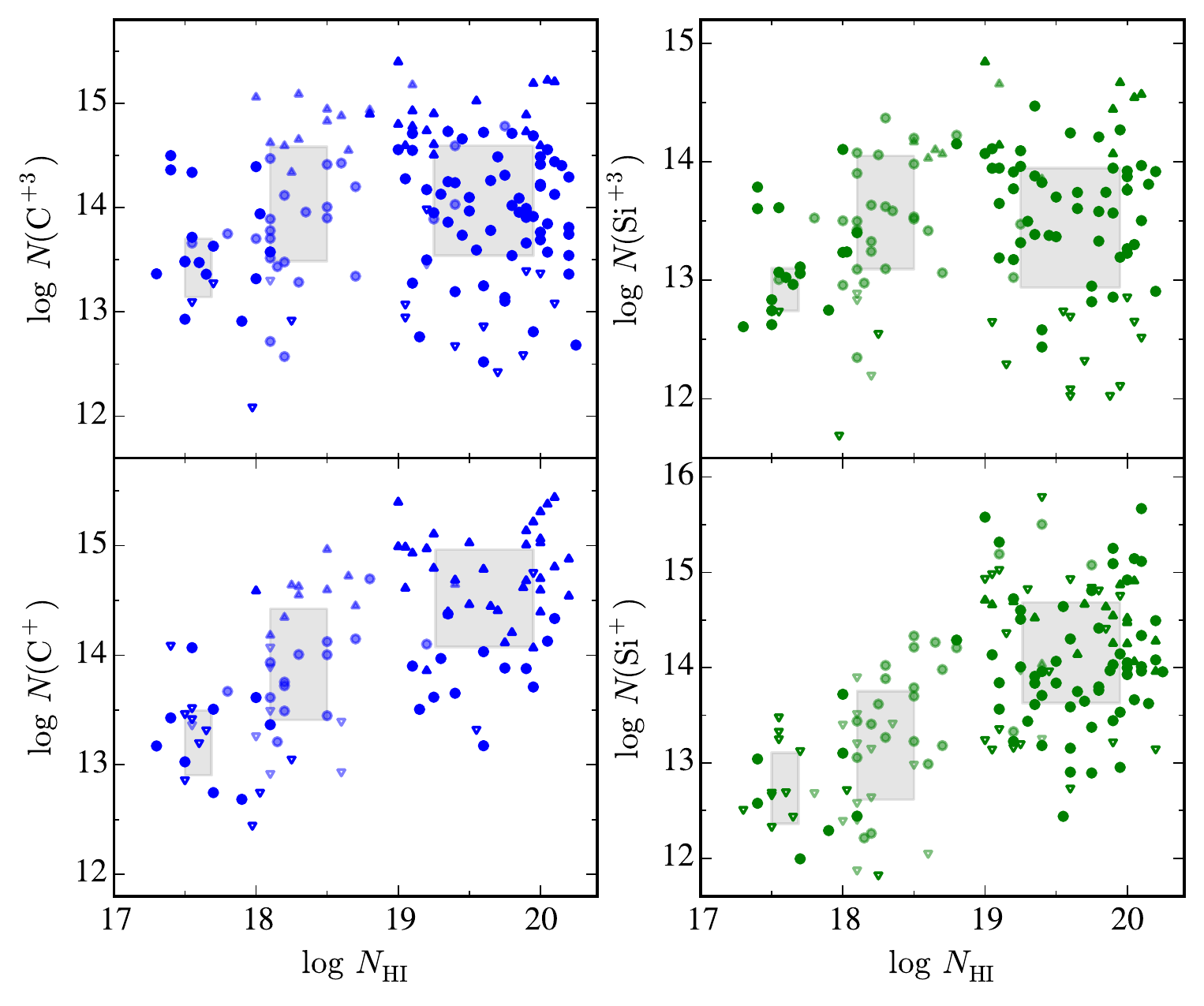}
\caption{Scatter plot of Si and C ionic column densities for the HD-LLS
  Sample.    Circles indicate measured values; their uncertainties are
  generally less than 0.1\,dex.  Triangles indicate limits to the
  values with the open symbols indicating upper limits.
  Lighter points mark LLSs with a poorly constrained
  \nhi\ value.   
  Gray boxes encompass 50-percent of the
  measurements in three logarithmic
  \nhi\ intervals: [17.3, 18.0), [18.0, 19.0), [19.0, 20.3).
  At all column densities, there is a large dispersion
  in the measurements.  Nevertheless, the low-ions (C$^+$, Si$^+$)
  exhibit a strong, positive correlation with \nhi\ value.  A Spearman
  rank test rules out the null hypothesis at $>99.99\%$ c.l.
}
\label{fig:SiC_scatter}
\end{figure}

\begin{figure}
\includegraphics[width=3.5in]{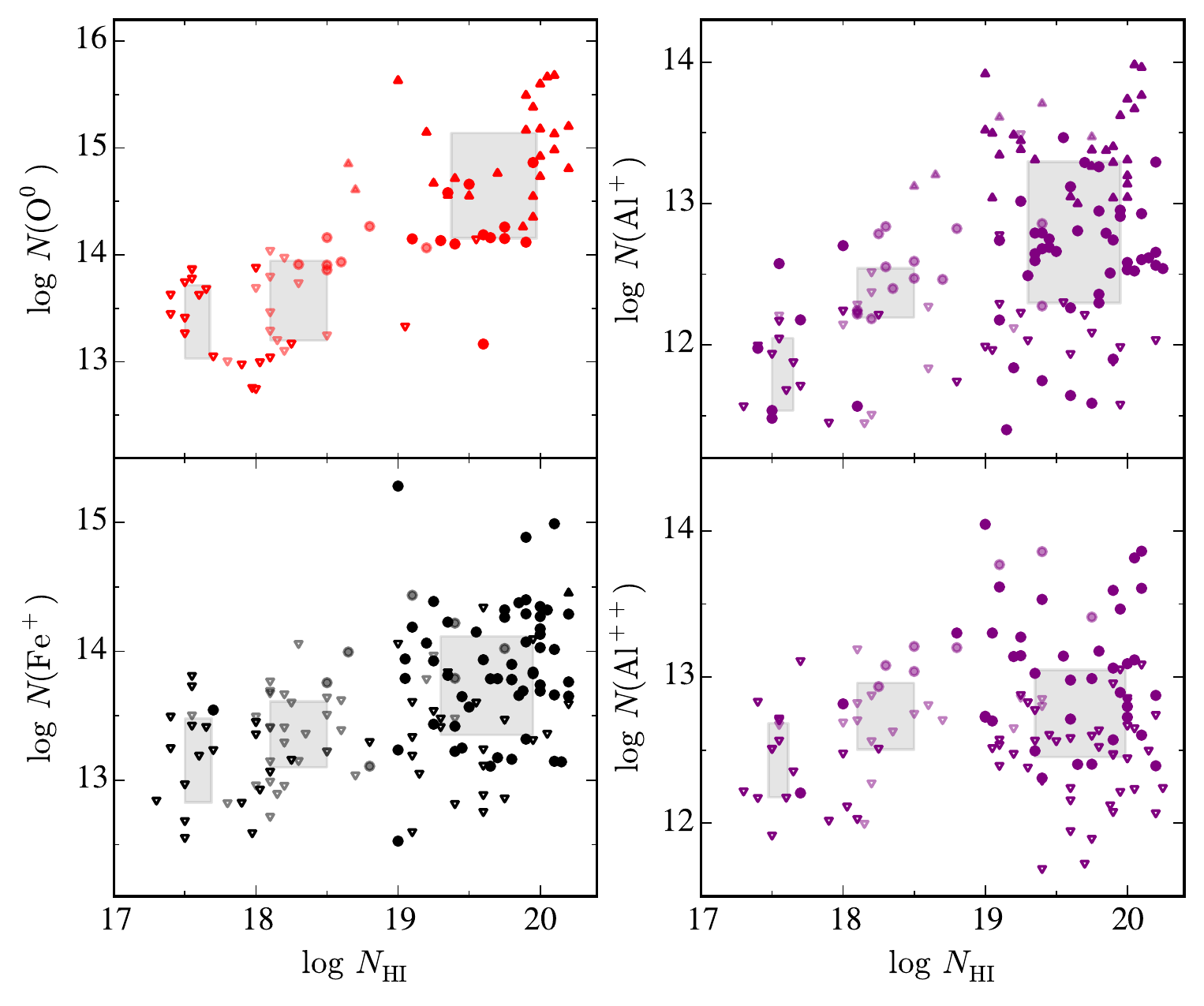}
\caption{Same as Figure~\ref{fig:SiC_scatter} but for four additional ions.
}
\label{fig:other_scatter}
\end{figure}

A complete set of tables and figures for the metal-line transitions
analyzed for each LLS are
given online. Tables~\ref{tab:SiCions} and
\ref{tab:other_ions} summarize the results for
Al$^+$, Al$^{++}$, Fe$^+$, 
C$^+$, C$^{+3}$, O$^0$, Si$^+$, and Si$^{+3}$.
A listing of all the measurements from this manuscript is provided in
the Appendix.
Figures~\ref{fig:SiC_scatter} and \ref{fig:other_scatter} show the column
density distributions for a set of Al, Fe, Si, C, and O atoms/ions as a function of the
LLS \nhi\ value.  Not surprisingly, the lower ionization states show
an obvious correlation\footnote{Taking limits as values, all of these
  ions have a Spearman rank test probability of less than 0.0001.} 
with \ion{H}{1} column density although there
is a large scatter at all values.  The near absence of positive
detections for \ion{O}{1} (i.e.\ $\N{O^0} < 10^{14} \cm{-2}$) at low
\nhi\ is also notable.  This suggests a rarity of high metallicity gas
in systems with $\mtll < 10$.
The high ions are also positively correlated with the neutral column
but with yet larger scatter and much smaller correlation coefficients.

\section{Results}
\label{sec:results}

In the following, we present a set of results derived from the column
density measurements of the previous sections.  For this manuscript, we 
restrict the analysis to an empirical investigation.  Future studies
will introduce additional models and analysis (e.g.\ photoionization
modeling) to interpret the data.
We further restrict the discussion to ionic abundances and defer the
analysis of kinematics to future work.

\begin{figure}
\includegraphics[width=3.5in]{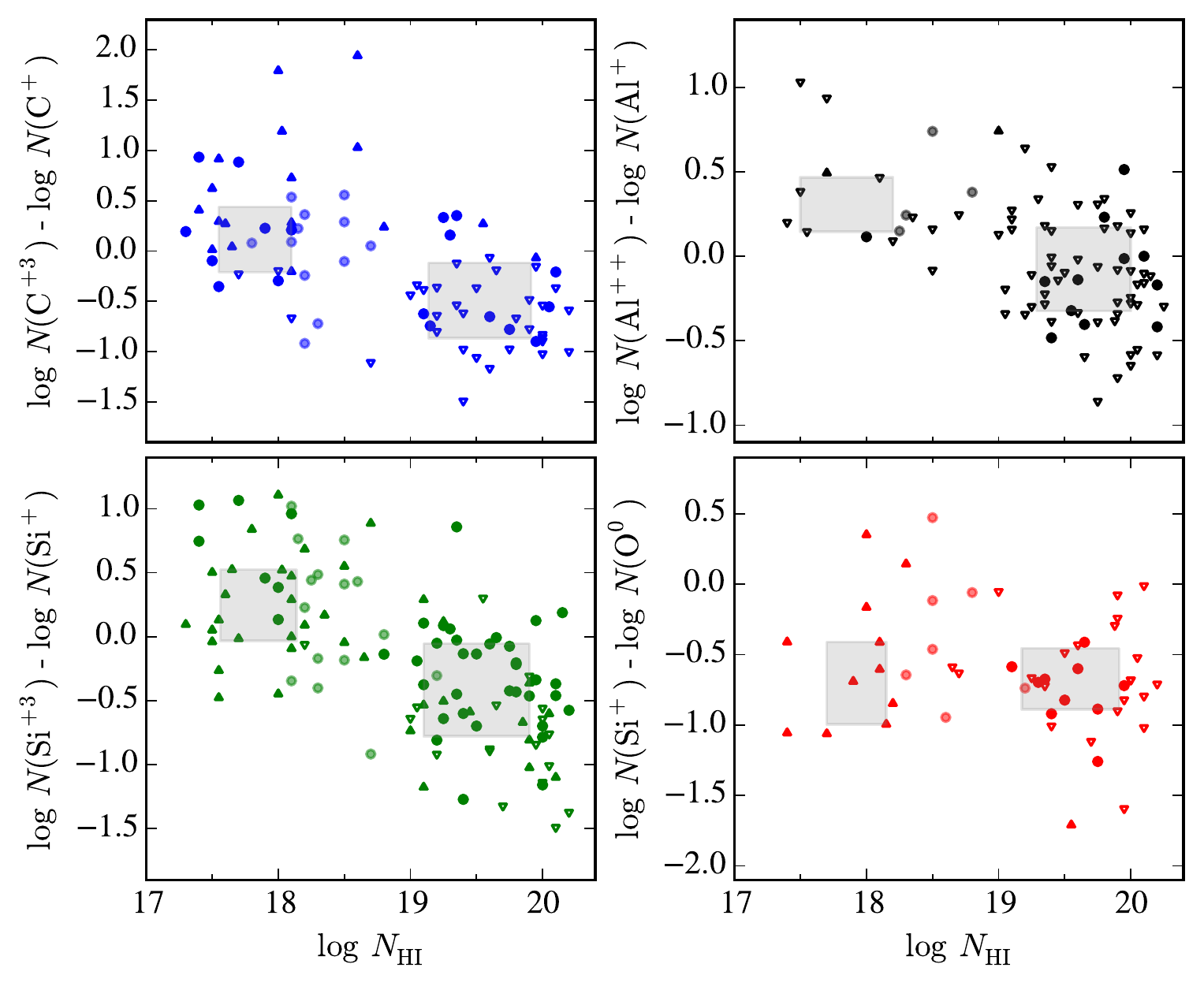}
\caption{Scatter plots of four ionic ratios that diagnose the
  ionization state of the LLSs.  
  Gray boxes encompass 50-percent of the
  measurements in two  logarithmic
  \nhi\ intervals: [17.3, 19.0), [19.0, 20.3).
  The C, Si, and Al ratios show strong
  evidence that the gas is more highly ionized at low \nhi\ values.
  Similarly, the set of Si$^+$/O$^0$ values exceeding $-0.7$\,dex are
  indicative of a highly ionized gas.  We further emphasize that the 
  $\log({\rm Si^+}/{\rm Si^{+3}}) \approx -0.5$\,dex values at high
    \nhi\ suggests that this gas is also partially ionized.
}
\label{fig:ionization}
\end{figure}

\subsection{Ionization State}

As noted above, a full treatment of the ionization state of the gas
including the comparison to models will be presented in a future
manuscript.  We may, however, explore the state of the gas empirically
through the examination of ionic ratios that are sensitive to the
ionization state of the gas.  Figure~\ref{fig:ionization}
presents four such ionic ratios against \nhi.  These primarily compare
ions of the same element (e.g.\ C$^{+3}$/C$^+$) to eliminate offsets
due to differing intrinsic chemical abundances (i.e.\ varying
abundance ratios).  
In this analysis, we have taken the
integrated column density across the entire LLS.  While there is
evidence for variations in these ratios within individual components,
these tend to be small
\citep[e.g.][Figure~\ref{fig:ionic_SLLS}]{ppo+10}. 
Therefore, the trends apparent in
Figure~\ref{fig:ionization} reflect the gross properties of the LLS
sample. 

All of the C$^{+3}$/C$^+$, Si$^{+3}$/Si$^+$, and Al$^{++}$/Al$^+$
ratios exhibit a strong anti-correlation with \nhi\ indicating an
increasing neutral fraction with increasing \ion{H}{1} opacity.
Taking limits as values, the Spearman rank test yields a probability
of less than $10^{-3}$ for the null hypothesis, in each case.
For all of these ions, the upper ionization state is dominant for 
$\mnhi \lesssim 10^{18.5} \cm{-2}$ and vice-versa for higher \nhi\ values.
We emphasize, however, that even at $\mnhi \approx 10^{20} \cm{-2}$
the observed ratios are frequently large, 
e.g.\ $\log({\rm Si^{+3}/Si^+}) \approx -0.5$\,dex.  
This suggests that the gas is predominantly ionized even
at these larger total \ion{H}{1} opacities.

This inference is further supported by the Si$^+$/O$^0$. 
Ignoring differential depletion, which we expect to be modest in LLSs,
the Si$^+$/O$^0$ ratio should trend towards the solar abundance ratio
($\epsilon_{\rm Si}/\epsilon_{\rm O} = - 1.2$\,dex)
in a neutral gas
given that Si and O are both produced in massive stars and are
observed to trace each other in astrophysical systems (e.g.\ stellar
atmospheres).
We identify, however, a significant sample of systems with $\mnhi
\approx 10^{18.5} \cm{-2}$ that have 
$\log({\rm Si^{+}/O^0})> -1$\,dex.
Because the majority of
ionization processes (e.g.\ photoionization, collisional ionization)
predict ${\rm Si^+/O^0} > $~Si/O \citep[e.g.][]{qpq3},
these measurements offer
further evidence that LLSs are highly ionized.

\begin{figure}
\includegraphics[width=3.5in]{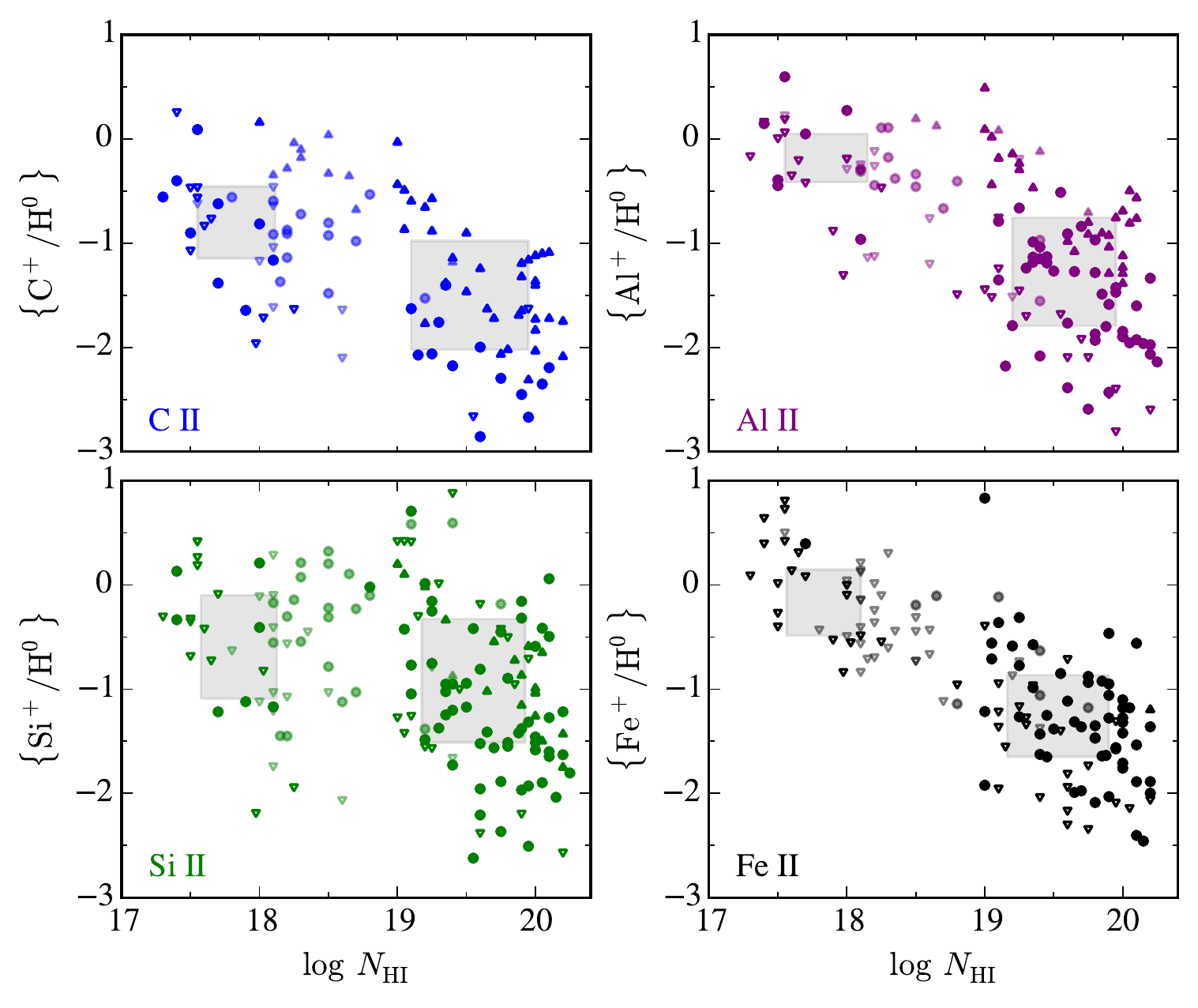}
\caption{Scatter plots of low-ion column densities relative to
  \ion{H}{1}, normalized to the solar abundance \psol{X}{i}{H}{0}\ and
  plotted against the LLS \nhi\ value.  
  Gray boxes encompass 50-percent of the
  measurements in two logarithmic
  \nhi\ intervals: [17.3, 19.0), [19.0, 20.3).
  If ionization corrections are small, \psol{X}{i}{H}{0}\ provides an
  estimate of the logarithmic metal abundance relative to solar.
  The measurements appear to indicate
  a declining trend of gas metallicity with increasing \nhi.  We
  argue, however, that this apparent trend is driven by ionization
  effects and the set of upper/lower limits at low/high \nhi\ values.
  Furthermore, given the large scatter at all \nhi\ values, it will be
  challenging to establish any trend between enrichment and \nhi\
  value in the LLSs.
}
\label{fig:low_MH}
\end{figure}

\begin{figure}
\includegraphics[width=3.5in]{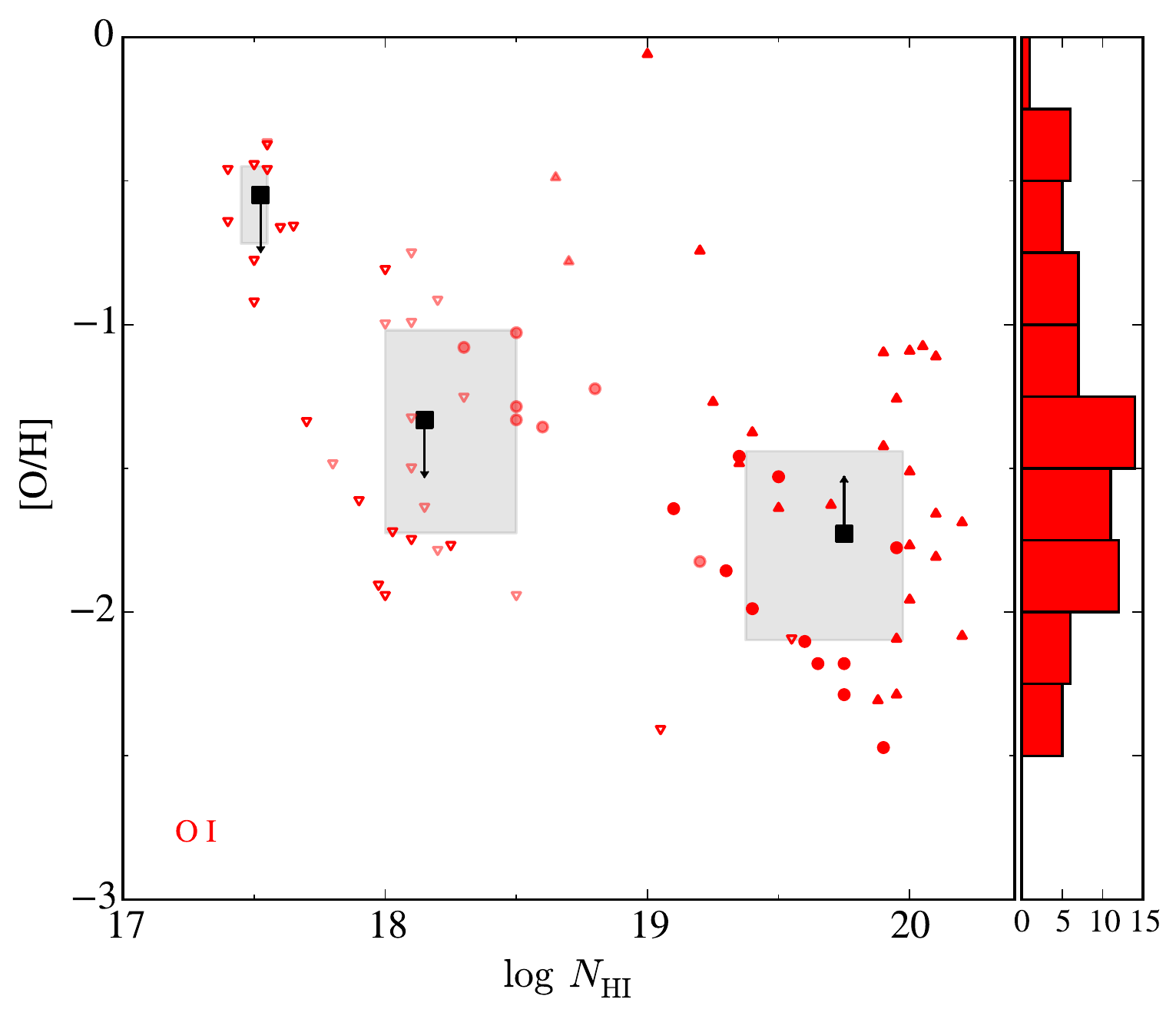}
\caption{Estimations of the oxygen metallicity in the LLSs where we
  have assumed that [O/H]=\psol{O}{0}{H}{0}, i.e.\ that ionization
  corrections are small for this ionic ratio (see the text).  
  Gray boxes encompass 50-percent of the
  measurements in three logarithmic
  \nhi\ intervals: [17.3, 18.0), [18.0, 19.0), [19.0, 20.3).
  Similar
  to the results for other low-ions (Figure~\ref{fig:low_MH}), there
  is an apparent decline in [O/H] with \nhi. This apparent trend,
  however, is primarily driven by the high incidence of upper/lower
  limits at low/high \nhi\ values.  In fact, the underlying trend may
  well be the opposite.  Compare the set of low [O/H] values and upper
  limits at $\mnhi \approx 10^{18} \cm{-2}$ with the large set of
  lower limits at $\mnhi \approx 10^{19.7} \cm{-2}$.
}
\label{fig:OH}
\end{figure}

\subsection{Metallicity}

A principal diagnostic of the LLSs is the gas metallicity, i.e.\ the
enrichment of the gas in heavy elements.
This quantity is generally characterized relative to the
chemical abundances observed for the Sun.  
For the following, we adopt the solar abundance scale compiled by
\cite{asplund09}, taking meteoritic values when possible.

Because the LLSs are significantly
ionized, the observed ionic abundances reflect only a fraction of the
total abundances of Si, O, H, etc.  
Therefore, a full treatment requires ionization modeling.  
We may, however, offer insight into the
problem by examining several ions relative to H$^0$.  To minimize
ionization corrections, one restricts the analysis to ionization
states dominant in a highly optically thick (i.e.\ neutral) medium.

The results for four low-ions are presented in
Figure~\ref{fig:low_MH}, normalized to the solar abundance.  
We have introduced here a new quantity and notation:
\psol{X}{i}{Y}{j}~$\equiv \log(\N{X^i}/\N{Y^j}) - \epsilon_{\rm X} +
\epsilon_{\rm Y}$, where $\epsilon_{\rm X}$ is the solar abundance on
the logarithmic scale for element X.  
This quantity explicitly ignores ionization corrections and should
not be considered a proper estimate of the chemical abundance ratio,
traditionally expressed as [X/Y].
In the cases where ionization corrections are negligible, however,
\psol{X}{i}{H}{0} = [X/H] and this quantity represents the
metallicity.  

A cursory inspection of the plots suggests a significant decline in 
metal content with increasing \nhi.  This apparent anti-correlation,
however, is driven by at least two factors.  First, larger \nhi\
implies larger metal column densities such that the transitions saturate
yielding a preponderance of lower limits.  By the same token, at low
\nhi\ values the transitions are often undetected yielding
upper limits to the ionic ratios.  Second, we have argued from
Figure~\ref{fig:ionization} that the gas is
increasingly ionized with decreasing \nhi.  For Si$^+$, C$^+$, and
Al$^+$, the ionization corrections for \psol{X}{i}{H}{0}\
are likely negative 
\citep[e.g.][]{pro99,fop11}, and
would lower the metallicity one infers from such ratios.
We believe these factors dominate the trends
apparent in Figure~\ref{fig:low_MH}.  

In fact, it is even possible that the true distribution exhibits the
opposite trend.  
Figure~\ref{fig:OH} shows [O/H] against \nhi\ for the LLSs where we
have assumed no ionization corrections, i.e.\ [O/H] =
\psol{O}{0}{H}{0}.
This approximation is justified by the fact that O$^0$ and H$^0$ have
very similar ionization potentials and their neutral states
are coupled by charge-exchange reactions.  This assumption may break down
at low \nhi\ values in the presence of a hard radiation field
\citep{sj98,ppo+10}, but the corrections are still likely to be
modest (several tenths dex).  Unfortunately, the measurements are
dominated by limits:  upper limits at $\mnhi < 10^{18.5} \cm{-2}$ and
lower limits at $\mnhi > 10^{19} \cm{-2}$.  Nevertheless, the data
require that [O/H]~$> -1.7$ for the SLLSs and indicate
[O/H]~$<-1.3$\,dex for LLSs with $\mnhi \approx 10^{18} \cm{-2}$.
We tentatively infer that the median metallicity is approximately
flat with \nhi\ and possibly increasing;  more strictly, we rule out
a steeply declining O/H metallicity with increasing \nhi. 
%
%
A similar conclusion may be drawn from the \psol{Si}{+}{H}{0}\ measurements
which scatter less from line-saturation.  The LLSs with $\mnhi \approx
10^{18} \cm{-2}$ show very few positive detections and have a median 
\psol{Si}{+}{H}{0}~$< -1$\,dex.  In contrast, the LLSs at
$\mnhi \approx 10^{19.5} \cm{-2}$ frequently exhibit
\psol{Si}{+}{H}{0}~$> -1$\,dex.

\begin{figure}
\includegraphics[width=3.5in]{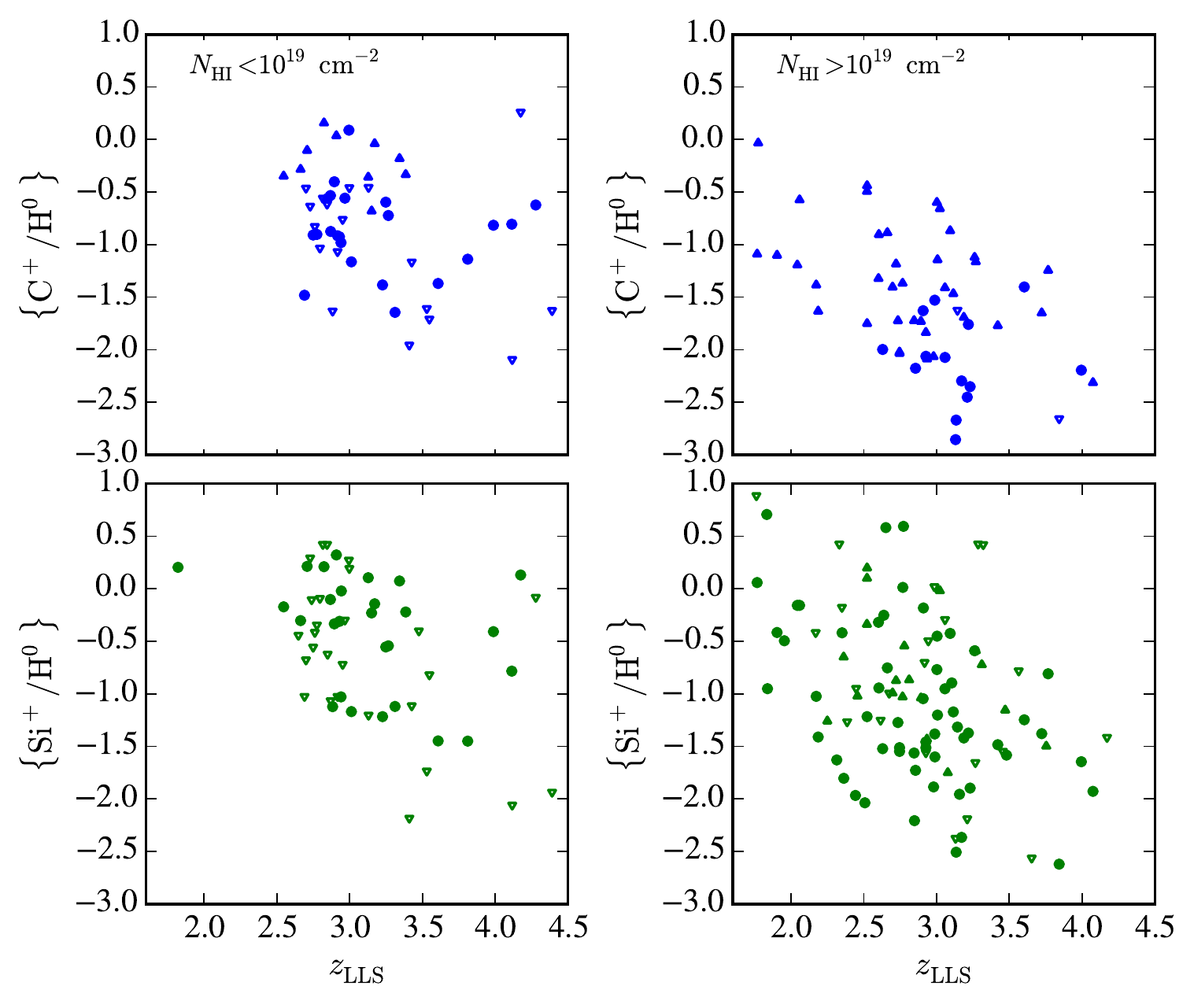}
\caption{Comparison of \psol{X}{i}{H}{0} measurements for 
C$^+$ and Si$^+$ against absorption redshift.  Left-hand panels are
for the LLSs with $\mnhi < 10^{19} \cm{-2}$ and the right-hand panels
are for the SLLS population.  All of the data show evidence for a
declining enrichment with increasing redshift, although this assertion
is statistically significant ($>99\%$ c.l.) only for the SLLS sample.
The absence of low values at $z \approx 2$ implies a reduced
incidence of near-pristine gas
with high \ion{H}{1} columns at that epoch.
}
\label{fig:z_metal}
\end{figure}

Another result apparent from Figure~\ref{fig:low_MH} is the
large dispersion in measurements at every \nhi\ value.  This is most
notable for Si$^+$ which has multiple transitions that
permit measurements of the column density 
over a larger dynamic range.  
At the largest \nhi\ values, the values/limits of 
\psol{Si}{+}{H}{0} span nearly four orders of magnitude! 
And although the measurements for LLSs with $\mnhi \approx 10^{17.5} -
10^{19} \cm{-2}$ include many upper limits, one
identifies values and upper limits with 
\psol{Si}{+}{H}{0}~$> -0.5$\,dex together with upper limits having
\psol{Si}{+}{H}{0}~$< -2$.
Clearly, any
underlying trend of enrichment with \nhi\ will be diluted
by the large intrinsic scatter within the LLSs.
One may even argue that if such a dispersion is indicative of multiple
astrophysical systems, then defining a mean of the LLS population has
limited scientific value.

Despite the large dispersion, we emphasize that very few of the LLS in
the HD-LLS Sample
are ``metal-free'', i.e.\ exhibiting no metal-line absorption and
therefore consistent with primordial abundances.
Of the \nlls\ LLSs, only 25 have no low-ion detections outside the
\lya\ forest and 18 of these exhibit a positive detection in a
higher-ion.   For the other 7, one has been previously been identified
as consistent with primordial \citep{fop11}.  The remainder have a
diversity of S/N and spectral coverage and therefore are generally
less sensitive to measuring a low metallicity.  Several will be
examined in greater detail in a future manuscript.  Nevertheless, we
may conclude that the incidence of very low metallicity gas ($<1/1000$
solar) is rare in the LLS population ($<5\%$).
Furthermore, none of the 82\,LLSs with $\mnhi > 10^{19.2} \cm{-2}$ are
metal-free\footnote{There is the possibility of a slight bias against
  our identifying metal-free SLLS but we have been as careful as
  possible to select systems based solely on the \lya\ profile.}.
By $z \sim 3$, gas that is dense enough to exhibit
a very high Lyman limit opacity has previously been 
polluted by heavy elements.  

At the opposite end of the enrichment distribution,
we identify 13~systems with a positive \psol{Si}{+}{H}{0}\ 
measurement that exceeds 0\,dex.   This includes four extreme examples
with \psol{Si}{+}{H}{0}~$> +0.5$\,dex.  
Because these four LLSs also have $\mnhi \ge 10^{19} \cm{-2}$ we expect
that corrections for ionization are modest \citep[see][]{poh+06} and that
these are truly super-solar abundances.
The others, however, have uncertainties consistent with
the gas being sub-solar even before accounting for ionization.
We conclude, subject to additional future analysis, 
that super-solar enrichment is also rare in the LLSs.

In Figure~\ref{fig:z_metal}, we examine \psol{Si}{+}{H}{0}\ and
\psol{C}{+}{H}{0}\ values as a function of redshift, splitting the
LLS sample at $\mnhi = 10^{19} \cm{-2}$.  The values for
the lower \nhi\ systems suggest a declining trend with increasing
redshift, e.g.\ in contrast to the lower redshift systems,
none of the $z>3.5$ LLSs have a positive detection of
\psol{C}{+}{H}{0}~$> -0.5$\,dex.
Even if we restrict analysis to positive detections, however, an
anti-correlation is not statistically significant.

Turning to the SLLS population, the \psol{X}{i}{H}{0}\ distributions
show obvious trends with redshift (limits not withstanding).
Treating all of the positive detections at their plotted values, a
Spearman's rank correlation test rules out the null hypothesis at
$>99\%$ c.l.   We interpret this anti-correlation as lower average
enrichment within the SLLS at higher redshift. This conclusion relies
on the assumption that ionization corrections will not evolve
significantly with redshift, which will be investigated in a future
work.
A similar decline in metallicity has been established in the DLA
population \citep[e.g.][]{pgw+03,rafelski+12} and has been interpeted as
resulting from the ongoing enrichment of galactic ISM with cosmic
time.  Future work will perform a quantitative comparison between the
two populations and explore the implications for the evolving
enrichment of optically thick gas at $z>2$.
In passing, we emphasize
the absence of low \psol{X}{i}{H}{0}\ values at $z \approx 2$ 
which implies a reduced
incidence of near-pristine gas
with high \ion{H}{1} columns at that epoch.

\begin{figure}
\includegraphics[width=3.5in]{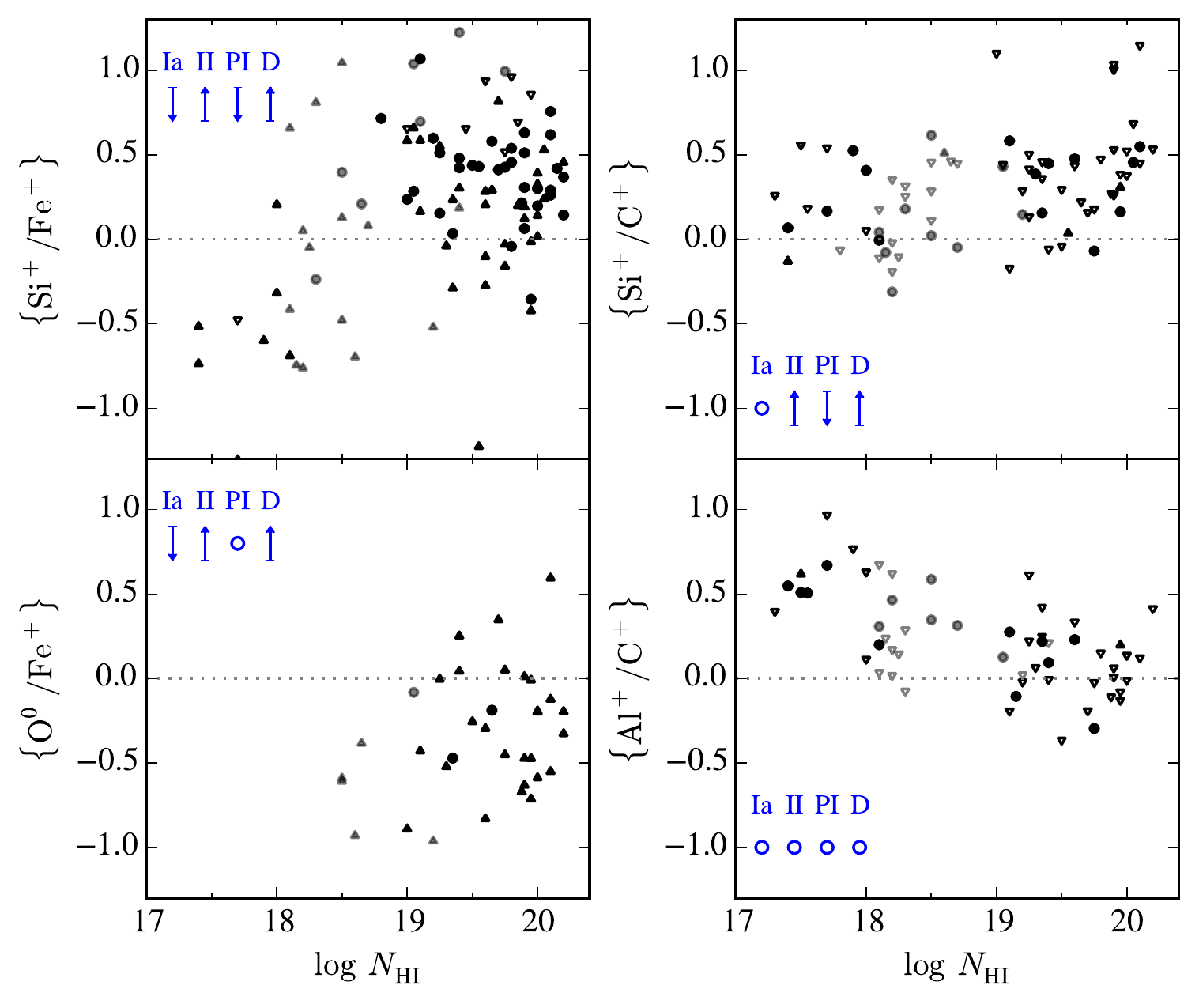}
\caption{Scatter plots of a series of ionic ratios, normalized to the
  solar relative abundances, against the \nhi\ values of the LLSs.
The left two panels show ratios related to the $\alpha$/Fe abundance.
Despite the preponderance of lower limits (especially for
\psol{O}{0}{Fe}{+}), and concerns on the ionization corrections, we
tentatively conclude that the LLSs exhibit super-solar $\alpha$/Fe
ratios, especially at large \nhi\ values.  The measurements in the
upper-right panel  indicate that Si/C is possibly enhanced by a few
0.1\,dex relative to solar although the majority of the sample is
consistent with [Si/C] = 0.
Similarly, the \psol{Al}{+}{C}{+}\ measurements are roughly consistent
with the solar relative abundance or possibly sub-solar.
In each panel, we indicate the expected offsets to the measurements
that would be due to Type~Ia (Ia) nucleosynthesis,
Type~II (II) nucleosynthesis,
photoionization (PI), 
and 
differential depletion (D).  
Circles indicate a small or unknown impact.
}
\label{fig:nucleo}
\end{figure}

\subsection{Nucleosynthetic Patterns}

It is the conventional wisdom that LLSs primarily trace gas outside
of the ISM of galaxies, e.g.\ within their dark matter halos (aka CGM) or 
at yet greater distances \citep{fpk+11,qpq6}.  Despite their separation from
galaxies, we have demonstrated that the LLSs are generally enriched in
heavy elements and provided evidence that their metallicity frequently
reaches $\sim 1/10$ solar abundance.
Therefore, a non-negligible fraction of this optically thick medium 
has been processed
through the furnaces of a stellar interior and presumably was
transported from a galaxy via one more physical processes.
One plausible transport process is an explosive event, e.g.\ a supernovae that
expelled the gas shortly after enriching it.  In this case, the gas
may exhibit a distinct nucleosynthetic pattern from
those observed for galactic ISM, i.e.\ if the supernovae ejecta did not mix
prior to escaping the system.
Additionally, the LLSs may couple the metal production within galaxies to
the enrichment of the diffuse IGM
\citep[e.g.][]{ahs+01,schaye03,steidel+10}.
This motivates comparison of the abundances for
these two diffuse and ionized phases.

We may explore several ionic ratios that trace
different nucleosynthesis channels.  As with metallicity, one
must account for ionization effects when interpreting the
results. Figure~\ref{fig:nucleo} plots four pairs of ions from the
dataset, again represented as \psol{X}{i}{Y}{j}\ with ionization
corrections explicitly ignored.  The figure also indicates
the probable offsets to the ratios if ionization effects were
important, as estimated from photoionization calculations (e.g.\
Prochaska 1999).
Similarly, we indicate the likely offsets from differential depletion
and the dominant nucleosynthesis channels (Type~Ia and Type~II
enrichment). 

The left-hand panels show two measures of $\alpha$/Fe, a key
diagnostic of the relative contributions of Type~Ia and Type~II
SNe nucleosynthesis \citep{tinsley79}.  Unfortunately, the 
\psol{O}{0}{Fe}{+} ratios are dominated by lower limits due to the
saturation of \ion{O}{1}~1302 and the non-detection of \ion{Fe}{2}
transitions.  The values are nearly consistent with a solar abundance
although there are at least 
two systems with \psol{O}{0}{Fe}{+}~$> +0.3$\,dex 
suggesting an $\alpha$-enhanced gas.  
Correcting for photoionization effects would only strengthen this
conclusion. 
These two LLSs also
exhibit a low metallicity ([O/H]~$\approx -2$) such that their
chemical signature is very similar to that of metal-poor Galactic
stars \citep{mcw97}.

Turning to \psol{Si}{+}{Fe}{+}, the sample is dominated by
measurements exceeding the solar abundance.  This includes a
non-negligible set of measurements with \psol{Si}{+}{Fe}{+}~$> +
0.5$\,dex,  and one may speculate that this represents the
metal-enriched ejecta of Type~II SNe.  The \psol{Si}{+}{Fe}{+}\ ratio,
however, is likely to require 
an ionization correction to accurately estimate Si/Fe.
This could explain, in part, the
positive \psol{Si}{+}{Fe}{+}\ values in Figure~\ref{fig:nucleo}.  On
the other hand, the highest \psol{Si}{+}{Fe}{+}\ values occur in LLSs
with high \nhi\ values where one expects ionization effects to be
minimal\footnote{Such gas may also experience
  differential depletion, i.e.\ elevated Si/Fe ratios in the gas phase
  to the refractory nature of these elements
  \citep[e.g.][]{jenkins09}.  If the gas is predominantly ionized,
  however,  the depletion levels may be modest and this effect would
  be small.}.
We conclude, therefore, that at least a subset of the LLS population
exhibits
super-solar $\alpha$/Fe ratios indicative of Type~II enrichment, even
in higher metallicity gas.

Previous studies of gas in the IGM at $z \sim 2$ have reported an
enhanced Si/C abundance \citep{ask+04}.  This result was derived
statistically from the pixel optical depth method and is sensitive to
the assumed model of the extragalactic UV background \citep[EUVB; see
also][]{simcoe11}.
The results for LLSs offer a mixed picture (Figure~\ref{fig:nucleo}).
There are a handful of positive \psol{Si}{+}{C}{+}\ values
up to $+0.5$\,dex, with the highest measurements at low metallicity.
On the other hand, the sample is dominated by upper limits 
(from \ion{C}{2}~1334 saturation) and over half of these have
\psol{Si}{+}{C}{+}~$<+0.3$\,dex.
Once again, photoionization corrections would only strengthen this
result. 
As such, the LLS observations do not appear to exhibit a 
high enrichment of Si/C than that previously
inferred for the IGM. 

Figure~\ref{fig:nucleo} also presents the set of \psol{Al}{+}{C}{+}\
measurements that are not fully compromised by line-saturation.
These data are consistent with the lighter element ratios in LLSs
having solar relative abundances. 
The preponderance of upper limits, however, allows
that Al could be under-abundant relative to C.

\begin{figure}
\includegraphics[width=3.5in]{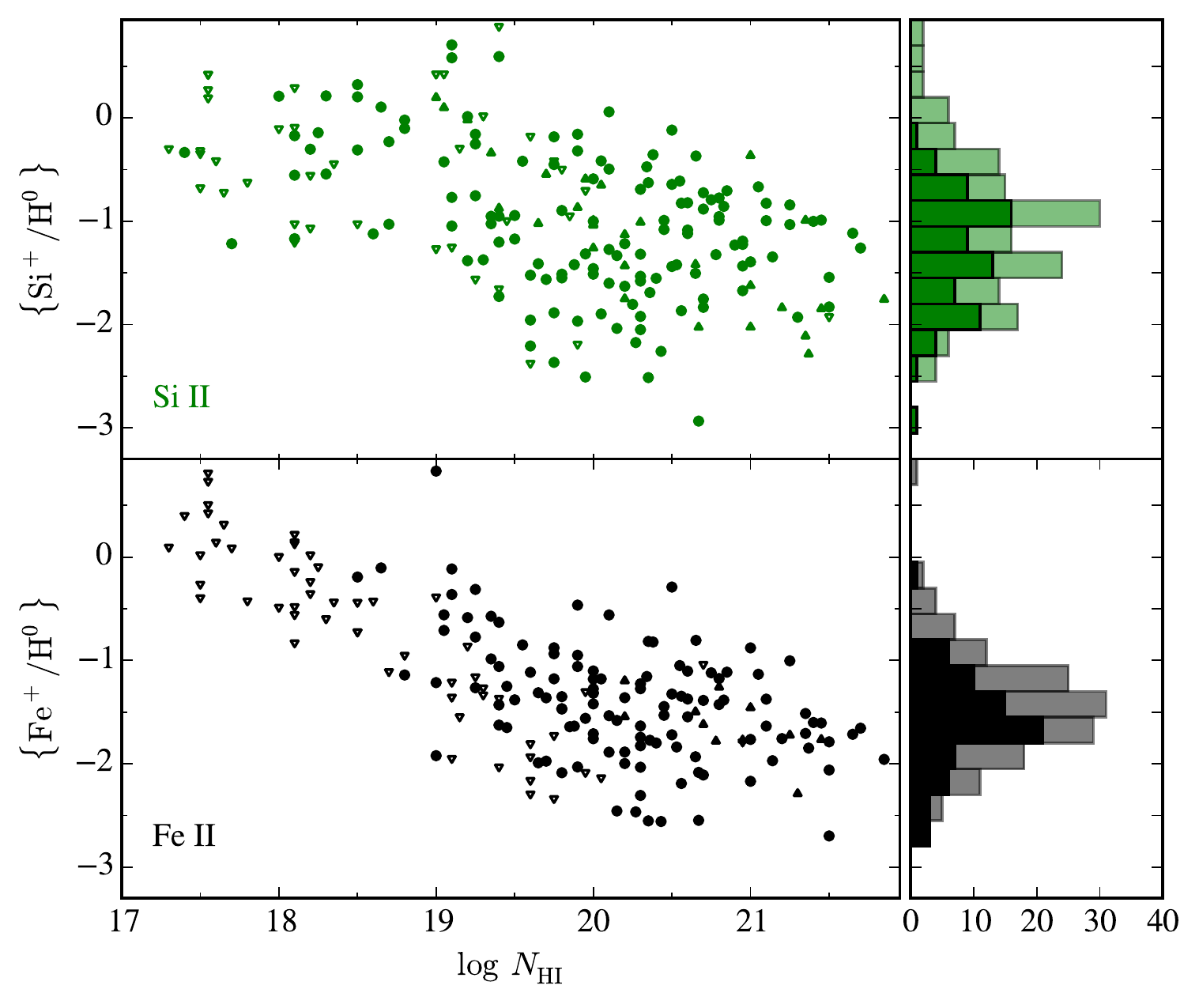}
\caption{
Comparison of the \psol{Si}{+}{H}{0}\ and \psol{Fe}{+}{H}{0}\
measurements for the LLSs and DLAs.  The latter are drawn from the
abundance compilation of \cite{rafelski+12}.  
For both ions, the DLAs show a continuous extension of the
measurements observed in the LLSs.  
The right hand panels compare the distributions of the DLAs (darker)
against those for the SLLSs (lighter), and emphasize the commonality 
between the two datasets.  In each case, we have treated upper 
and lower limit estimates as values.
}
\label{fig:XH_LLS_DLA}
\end{figure}

\subsection{Comparisons}

We have restricted the HD-LLS Sample to systems with $\mnhi < 10^{20.3}
\cm{-2}$ to exclude the DLAs.  This was partly motivated by the
expectation that the majority of LLSs are predominantly ionized and therefore
physically distinct from the neutral gas comprising DLAs.
It was also motivated by the desire to examine this optically thick
gas separately from the decades of research on the DLAs.
Nonetheless, the $\mnhi = 10^{20.3} \cm{-2}$ criterion is primarily an
observationally defined boundary and one may gain insight into the
nature of the LLSs through a combined comparison.
Such analysis has been performed previously for the SLLS by 
\cite{peroux03,som13}.

We consider two such comparisons here.  Figure~\ref{fig:XH_LLS_DLA}
presents the \psol{Si}{+}{H}{0}\ and \psol{Fe}{+}{H}{0}\ measurements for
the HD-LLS Sample together with measurements from the sample of
DLAs of \cite{rafelski+12}.  For both datasets, we have restricted to 
$z_{\rm abs} = [1.6,3.3]$ to minimize trends related to redshift
evolution.
To zeroth order, the DLA measurements extend in a roughly continuous
manner from the 
measurements of the LLSs.  Indeed, comparing the samples of DLA
measurements with the SLLSs (taking limits at their values),
one observes overlapping distributions with similar median values.
The only notable distinction, perhaps, is the small set of LLSs with 
$\mnhi \approx 10^{19} \cm{-2}$ and high \psol{X}{i}{H}{0}\ values
(exceeding 0\,dex for Si$^+$).
This suggests a higher incidence of highly enriched gas in the LLS,
although we caution it could be partly an effect of ionization.
The dispersion in the measurements is also larger for the LLSs, and
is likely higher than suggested by the Figure given the preponderance
of upper/lower limits for the LLS/DLA.

\begin{figure}
\includegraphics[width=3.5in]{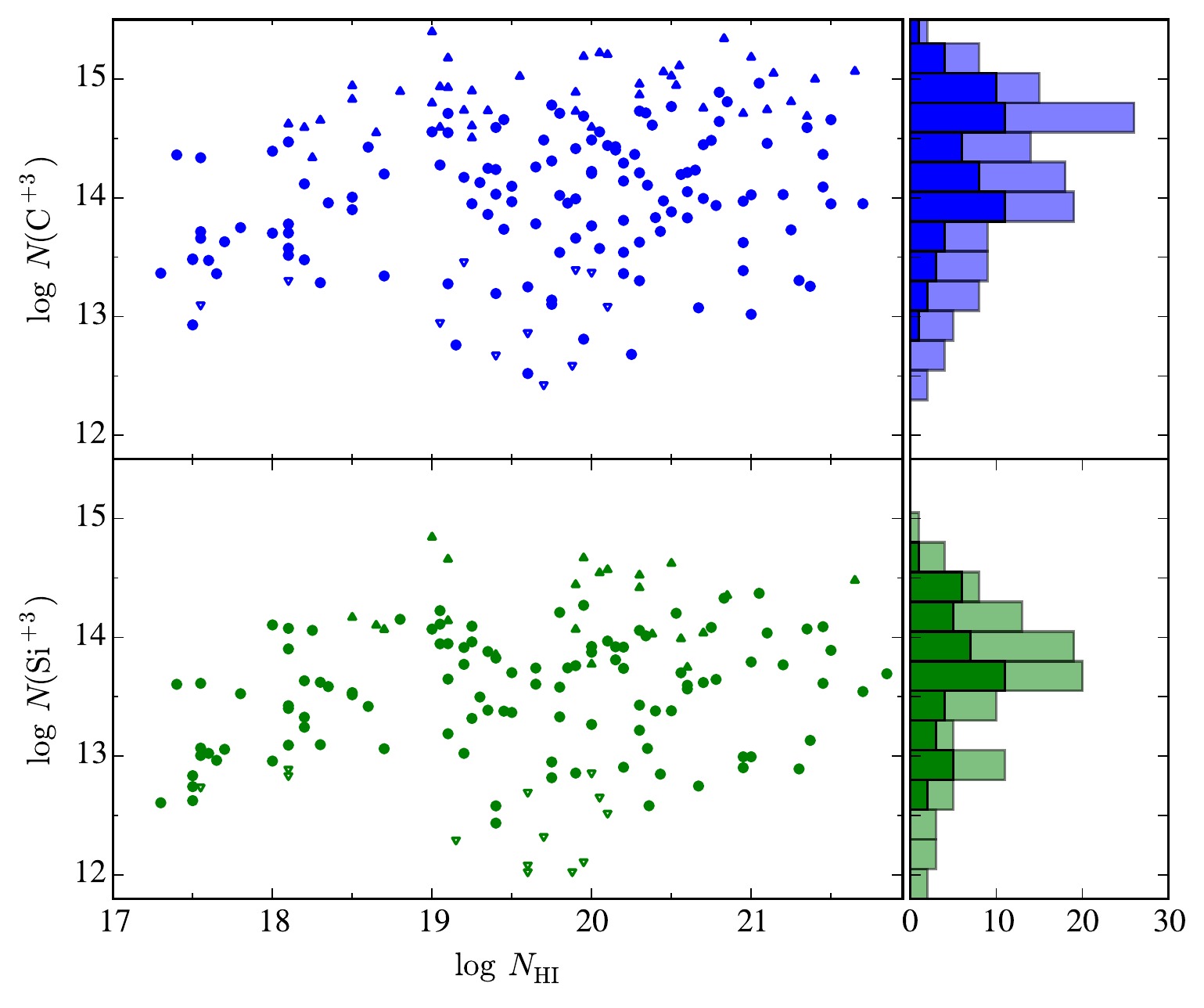}
\caption{
Comparison of the high-ion column densities measured for the LLS and a
representative set of DLAs \citep[drawn from][]{rafelski+12,marcel13}.
Similar to the low-ion abundances, the measurements show a continuous
transition from the LLS regime to higher \nhi\ values.  
As such, there is substantial overlap in the distribution of
measurements and limits (right-hand panels compare the values for DLAs
[darker] and SLLSs [lighter]).
One notable and surprising difference is that the SLLSs show a higher
incidence of low columns of C$^{+3}$ and Si$^{+3}$.
Despite tracing predominantly neutral gas, the DLAs also mark 
a reservoir of highly ionized gas that frequently exceeds the medium
encompassing the LLSs.
}
\label{fig:High_LLS_DLA}
\end{figure}

Turning to the higher ionization states, Figure~\ref{fig:High_LLS_DLA}
presents the C$^{+3}$ and Si$^{+3}$ column densities from the 
LLSs and DLAs.  
Once again, the DLA distribution extends in a nearly continuous manner
from the upper end of the LLS data and the column density
distributions for the SLLSs and DLAs are similar.
Together, Figures~\ref{fig:XH_LLS_DLA} and \ref{fig:High_LLS_DLA} lend
support to scenarios that envision LLSs as the outer layers 
of gas surrounding DLAs, i.e.\ these systems frequently sample the same
structures.   Such physical associations may be examined by studying
DLAs and LLSs along pairs of quasar sightlines, an active area of
research \citep{ehm+07,fhp+14,rubin+15}.

Examining the high-ion comparison further, there is at least 
one import distinction: the LLSs and
especially the SLLSs show a much higher incidence of low 
$\N{C^{+3}}$ and $\N{Si^{+3}}$ values.  This is unexpected given that 
(i) the LLSs trace highly ionized gas;
(ii) the DLAs trace predominantly neutral gas that is physically
distinct from the high-ions \citep{wp00a,pcw+08}.
The results presented here indicate that the gas layers giving rise to
DLAs are embedded in a reservoir of highly ionized gas that frequently
exceeds the typical surface density in LLSs.
This follows previous work which has inferred high quantities
of both neutral and ionized gas for the DLAs
\citep{fpl+07,lehner+14}.  
It further suggests that high-ions more closely trace higher density
regions in the universe and/or may reflect a difference in the masses
of the dark matter halos hosting LLSs and DLAs.

Lastly, we have compared our measurements against the small set of
literature values for SLLSs at $z>2$ \citep{mirka03,som13}.
The low-ion column densities are typically larger in those
publications, consistent with the higher \nhi\ values of the SLLSs
that were sampled.

\section{Summary}
\label{sec:summary}

We have constructed a sample of \nhdlls\ LLSs at $z \sim 2-4$ observed
at high-dispersion with spectrometers on the Keck and Magellan
telescopes which constitute 
the HD-LLS Sample.
In this manuscript, we present the
complete sample and present column density measurements of \ion{H}{1}
and associated metal absorption.  For the latter, analysis was
restricted to transitions redward of the \lya\ forest and has focused
on commonly detected species.  These measurements and the associated
spectra are made available online with this 
publication\footnote{http://www.ucolick.org/$~\sim$xavier/LLS}.
This constitutes, by roughly an order of magnitude, the largest high
redshift sample of LLS analyzed in this manner.

We have explored empirical trends in the column density measurements
and report statistically significant ($>99.99\%$) correlations between
the low-ion (e.g.\ Si$^+$, C$^+$) columns and \nhi.
High-ion species (Si$^{+3}$, C$^{+3}$) are detected in nearly all LLSs
and their column densities also correlated with \nhi.  
Examining ionic ratios sensitive to the ionization state
(e.g.\ C$^{+3}$/C$^+$, Si$^{+3}$/Si$^+$), we conclude that the LLSs
are predominantly ionized with more highly ionized gas in lower \nhi\
systems.

Ratios of low-ion column densities to \nhi\ indicate a wide spread in
metal-enrichment within the LLSs, likely spanning four orders of
magnitude.  Only a small subset ($\lesssim 5\%$) of the HD-LLS Sample have
no positive detections of associated metals, consistent with
primordial abundances.  None of the LLSs with $\mnhi \ge 10^{19.2}
\cm{-2}$ are `metal-free'.  We conclude that a very high percentage of
high-density gas at $z \sim 3$ was previously enriched to $\gtrsim
1/1000$ solar abundance.  The HD-LLS Sample also exhibits a small
subset ($\sim 10\%$) of LLSs that have solar or super-solar
enrichment.  These likely represent the most enriched gas reservoirs
in the high redshift universe.

Lastly, we have examined several ionic ratios that are sensitive to
the nucleosynthetic enrichment history of the gas.  The preponderance
of elevated Si$^+$/Fe$^+$ and O$^0$/Fe$^+$ measurements suggest the
LLSs have an $\alpha$-enhancement characteristic of Type~II
nucleosynthesis.  In contrast, the Si$^+$/C$^+$ and Al$^+$/C$^+$
ratios are consistent with solar relative abundances.

Future manuscripts on the HD-LLS Sample will:
 (i) study the metallicity distribution of the LLSs accounting for
 ionization effects and will estimate the contribution of optically
 thick gas to the cosmic metal budget;
 (ii) examine the kinematic characteristics to constrain the physical
 origin of the gas;
 (iii) offer constraints on the \nhi\ frequency distribution for
 optically thick gas.

\acknowledgments

J. X. P. was supported by NSF grants AST-1010004 and AST-1412981.
MF acknowledges support by the Science and Technology Facilities
Council, grant number ST/L00075X/1. 
We thank Claude-Andr\'e Faucher-Giguerre for kindly providing his
continuum fits to MIKE spectra.
We acknowledge the contributions of Wal Sargent and Brian Penprase in
collecting a portion of the ESI data and Arthur M. Wolfe, Marcel
Neeleman, and Marc Rafelski for their
contributions to the Keck observations.

Much of the data presented herein were obtained at the W.M. Keck
Observatory, which is operated as a scientific partnership among the
California Institute of Technology, the University of California, and
the National Aeronautics and Space Administration. The Observatory was
made possible by the generous financial support of the W.M. Keck
Foundation.  Some of the Keck data were obtained through the NSF
Telescope System Instrumentation Program (TSIP), supported by AURA
through the NSF under AURA Cooperative Agreement AST 01-32798 as
amended.

\clearpage

\appendix

{\bf APPENDIX: Measurements for Individual LLS}

\begin{figure}
\figurenum{16}
\includegraphics[width=6in]{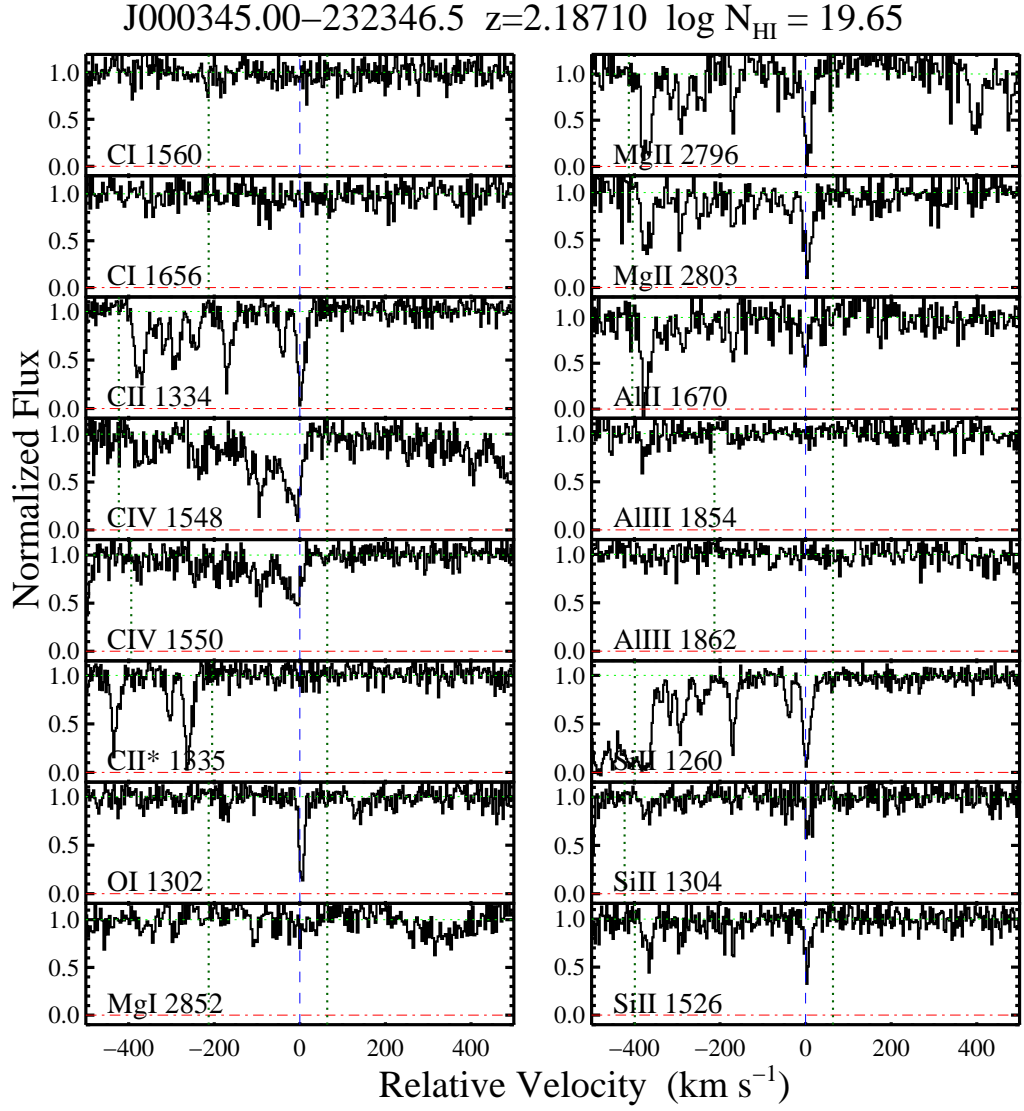}
\caption{Velocity plots for the HD-LLS Sample}
\label{fig:vel_plot_ex}
\end{figure}

Table~\ref{tab:all_lines} lists measurements for all of the
metal-line transitions analyzed in this manuscript and Figures
showing velocity plots are provided in the on-line materials
(Figure~\ref{fig:vel_plot_ex} shows one example).
The analysis was restricted to lines outside the \lya\ forest and those
lines that are not
severely blended with another feature or compromised by 
sky-subtraction residuals.






\begin{deluxetable*}{lccccccccccccc}
\tablewidth{0pc}
\tablecaption{IONIC COLUMN DENSITIES \label{tab:all_lines}}
\tabletypesize{\tiny}
\tablehead{\colhead{Quasar} & \colhead{RA} & \colhead{DEC} & \colhead{$z_{\rm abs}$} 
& \colhead{\nhi} 
& \colhead{$\lambda$} & \colhead{$v_{\rm lim}$} & \colhead{flg} & \colhead{$N_\lambda$} 
& \colhead{$\sigma(N)$} & \colhead{Ion} & \colhead{flg} & \colhead{$N_{\rm ion}$} & 
\colhead{$\sigma(N)$} \\ 
 & (J2000) & (J2000) & & & (\AA) & (km/s) } 
\startdata 
Q0001-2340 & 00:03:45 & -23:23:46.5& 2.18710& 19.65& 1334.5323 & $-423,64$ & 2 & 14.45 & 99.99& 6,2 & 2 & 14.45 & 99.99\\ 
&&&&& 1335.7077 & $-205,64$ & 4 & 13.10 & 99.99\\ 
&&&&& 1548.1950 & $-423,64$ & 0 & 14.26 & 0.01& 6,4 & 1 & 14.26 & 0.05\\ 
&&&&& 1550.7700 & $-394,64$ & 0 & 14.25 & 0.02\\ 
&&&&& 1302.1685 & $-213,64$ & 0 & 14.16 & 0.04& 8,1 & 1 & 14.16 & 0.05\\ 
&&&&& 2852.9642 & $-213,64$ & 4 & 11.72 & 99.99& 12,1 & 3 & 11.72 & 99.99\\ 
&&&&& 2796.3520 & $-413,64$ & 2 & 13.43 & 99.99& 12,2 & 1 & 13.64 & 0.05\\ 
&&&&& 2803.5310 & $-404,64$ & 0 & 13.65 & 0.02\\ 
&&&&& 1670.7874 & $-405,64$ & 2 & 13.00 & 99.99& 13,2 & 2 & 13.00 & 99.99\\ 
&&&&& 1854.7164 & $-213,64$ & 4 & 12.40 & 99.99& 13,3 & 3 & 12.40 & 99.99\\ 
&&&&& 1862.7895 & $-213,64$ & 4 & 12.69 & 99.99\\ 
&&&&& 1260.4221 & $-399,64$ & 2 & 13.81 & 99.99& 14,2 & 1 & 13.75 & 0.05\\ 
&&&&& 1304.3702 & $-423,64$ & 0 & 13.55 & 0.09\\ 
&&&&& 1526.7066 & $-399,64$ & 0 & 13.83 & 0.03\\ 
&&&&& 1808.0130 & $-213,64$ & 4 & 14.86 & 99.99\\ 
&&&&& 1393.7550 & $-411,64$ & 0 & 13.78 & 0.01& 14,4 & 1 & 13.74 & 0.05\\ 
&&&&& 1402.7700 & $-421,64$ & 0 & 13.64 & 0.02\\ 
&&&&& 1250.5840 & $-79,64$ & 0 & 14.19 & 0.13& 16,2 & 1 & 14.19 & 0.13\\ 
&&&&& 1608.4511 & $-213,64$ & 4 & 13.49 & 99.99& 26,2 & 1 & 13.11 & 0.05\\ 
&&&&& 2344.2140 & $-213,64$ & 0 & 13.22 & 0.07\\ 
&&&&& 2374.4612 & $-213,64$ & 4 & 13.46 & 99.99\\ 
&&&&& 2382.7650 & $-213,64$ & 0 & 13.01 & 0.04\\ 
&&&&& 2586.6500 & $-213,64$ & 4 & 13.15 & 99.99\\ 
&&&&& 2600.1729 & $-328,64$ & 0 & 13.25 & 0.04\\ 
&&&&& 1317.2170 & $-213,64$ & 4 & 13.41 & 99.99& 28,2 & 3 & 13.40 & 99.99\\ 
&&&&& 1370.1310 & $-213,64$ & 4 & 13.40 & 99.99\\ 
&&&&& 1454.8420 & $-213,64$ & 4 & 13.56 & 99.99\\ 
&&&&& 1741.5531 & $-213,64$ & 4 & 13.60 & 99.99\\ 
&&&&& 1751.9157 & $-213,64$ & 4 & 13.83 & 99.99\\ 
&&&&& 2026.1360 & $-213,64$ & 4 & 12.38 & 99.99& 30,2 & 3 & 12.38 & 99.99\\ 
\hline 
PX0034+16 & 00:34:54.8 & +16:39:20& 3.75397& 20.05& 1548.1950 & $-242,65$ & 0 & 13.85 & 0.02& 6,4 & 1 & 13.85 & 0.05\\ 
&&&&& 1550.7700 & $-118,189$ & 2 & 13.68 & 99.99\\ 
&&&&& 1670.7874 & $-187,189$ & 0 & 12.52 & 0.04& 13,2 & 1 & 12.52 & 0.05\\ 
&&&&& 1854.7164 & $-115,120$ & 4 & 12.24 & 99.99& 13,3 & 3 & 12.24 & 99.99\\ 
&&&&& 1862.7895 & $-89,129$ & 4 & 12.57 & 99.99\\ 
&&&&& 1526.7066 & $-54,122$ & 2 & 14.06 & 99.99& 14,2 & 2 & 14.06 & 99.99\\ 
&&&&& 1808.0130 & $-133,138$ & 4 & 14.73 & 99.99\\ 
&&&&& 1393.7550 & $-160,133$ & 0 & 13.30 & 0.02& 14,4 & 1 & 13.30 & 0.05\\ 
&&&&& 1741.5531 & $-187,189$ & 4 & 14.29 & 99.99& 28,2 & 3 & 13.72 & 99.99\\ 
&&&&& 1751.9157 & $-86,189$ & 4 & 13.72 & 99.99\\ 
&&&&& 2026.1360 & $-187,189$ & 4 & 13.16 & 99.99& 30,2 & 3 & 13.16 & 99.99\\ 
\hline 
\enddata 
\tablecomments{Columns are as follows: 
(1) Quasar name; 
(2,3) RA/DEC; 
(4) Absorption redshift of LLS; 
(5) HI column Density; 
(6) Rest wavelength of transition; 
(7) Velocity limits (min/max) for integration relative to $z_{\rm abs}$; 
(8) Flag on individual measurement: [0,1=standard measurement; 2,3=Lower limit; 4,5=Upper limit]; 
(9) $\log_{10}$ column density; 
(10) Standard deviation on $\log_{10} N$.  Limits are given a value of 99.99; 
(11) Ion [atomic number, ionization state]; 
(12) Flag for the ionic column density [1=standard measurement; 2=Lower limit; 3=Upper limit]; 
(13) $\log_{10}$ column density for the ion; 
(14) Standard deviation on $\log_{10} N_{\rm ion}$.  Limits are given a value of 99.99; 
} 
\tablecomments{[The complete version of this table is in the 
electronic edition of the Journal.  
The printed edition contains only a sample.]} 
\end{deluxetable*} 

\clearpage

\figsetstart
\figsetnum{16}
\figsettitle{Velocity Plots for the Lyman Limit Systems}

\figsetgrpstart
\figsetgrpnum{16.1}
\figsetgrptitle{J000345.00-232346.5_z2.187}
\figsetplot{f16_1.eps}
\figsetgrpnote{Velocity plots for the HD-LLS Sample}
\figsetgrpend

\figsetend

\end{document}